\documentclass[12pt, preprint]{aastex}

\begin{document}

\title{Investigating on the Nuclear Obscuration in Two Types of
Seyfert 2 Galaxies}

\author{X. W. Shu, J. X. Wang, P. Jiang, L. L. Fan, and T. G. Wang}

\affil{Center for Astrophysics, University of Science and Technology of China, Hefei, Anhui 230026, P. R. China \\
Joint Institute of Galaxies and Cosmology, USTC and SHAO, CAS 
}

\email{jxw@ustc.edu.cn}

\begin{abstract}
We build a large sample of Seyfert 2 galaxies (Sy2s) with both optical 
spectropolarimetric and X-ray data available, in which 29 Sy2s with the
detection of polarized broad emission line (PBL) and 25 without. 
We find that for luminous Sy2s with L$_{[\rm O~III]}$ $>$ 10$^{41}$ erg 
s$^{-1}$, sources with PBL have smaller X-ray absorption column density
comparing with those without PBL (at 92.3\% confidence level):
most of the Sy2s with N$_{\rm H}<$10$^{23.8}$ cm$^{-2}$ show PBL (86\%, 12 out 14),
while the fraction is much smaller for sources with heavier obscuration
(54\%, 15 out 28).
The confidence level of the difference in absorption bounces up to 99.1\% while using the 
"T" ratio ($\rm F_{2-10~keV}/F_{[O~III]}$) as an indicator.
We rule out observation or selection bias as the origin for the difference.
Our results, for the first time with
high statistical confidence, show that,
in additional to the nuclei activity, 
the nuclear obscuration also plays an important role
in the visibility of PBL in Sy2s. 
These results can be interpreted in the framework of the unified model. 
We can reach these results in the unified model if: a) the absorption 
column density is higher at large inclinations and b) the scattering 
region is obscured at large inclinations.
\end{abstract}


\section{Introduction}
The AGN unification model proposes that Seyfert 1 and 2 galaxies (Sy1s and 
Sy2s hereafter) are intrinsically the same objects, and the absence of broad 
emission lines in Sy2s is ascribed to the obscuration along the line of sight
by a pc-scale dusty torus (see the review by Antonucci 1993). 
The most convincing evidence is the detection of polarized broad emission 
lines (hereafter PBL) in some Seyfert 2 galaxies (Antonucci \& Miller 1985; Tran 1995; Heisler, Lumsden \& Bailey 1997; Moran et al. 2000; Lumsden et al. 2001; Tran 2001). Similarly 
infrared (IR) observations showed the existence of obscured broad line regions 
(BLRs) in Sy2s (Veilleux, Goodrich \& Hill 1997). Further evidence supporting the 
unification model comes from X-ray observations of Sy2s which show large 
amounts of obscuration, typically above 10$^{23}$ cm$^{-2}$ (e.g., Turner 
et al. 1997; Bassani et al. 1999). 

Despite observations do generally support orientation based unification model, 
only $\sim$ 50\% of Sy2s show broad lines in the polarized spectrum (e.g. Tran
2001, Gu \& Huang 2002). With an optical spectropolarimetric study of a well-defined and 
statistically complete $IRAS$ 60-$\mu$m selected Sy2 sample, Heisler et al. 
(1997) found a relationship between the detectability of polarized
broad H$\alpha$ and the $IRAS$ $f_{60}/f_{25}$ flux ratio that only those
galaxies with warm $IRAS$ colors ($f_{60}/f_{25}$ $<$ 4.0) show PBL.
Heisler et al. suggested that the detectability of PBL simply depends on the
inclination of the torus to the line of sight: in a Sy2 with the torus
highly inclined, cooler infrared color is expected, and the broad-line 
scattering screen could also be obscured.

A simple prediction of the inclination-related model is that Sy2s without
PBL (hereafter NPBL Sy2s) should show higher absorption column density 
since they are more inclined than the Sy2s with PBL (hereafter PBL Sy2s).
However, following studies have claimed no difference in the absorption column 
density between two types of Sy2s (Alexander 2001; Tran 2001, 2003).
Furthermore, as Alexander (2001, also see Lumsden et al 2001; Tran 2001, 2003; 
Gu \& Huang 2002) pointed out, the difference in the $IRAS$ $f_{60}/f_{25}$ 
flux ratio is not
an good indicator of the inclination but the relative strength of galactic
and Seyfert emission.
These studies (also see Cheng et al. 2002; Lumsden \& Alexander 2001), instead, have shown that the presence of PBL in Sy2s depends on the AGN
luminosity: Sy2s with PBL have higher luminosity comparing with Sy2s
without PBL. Explanations to the observational results include:
a) The contribution from the host galaxy or from a
circumnuclear starburst would dilute the nuclear optical spectrum,
making the detection of PBL more difficult for Sy2s with lower luminosity
(Alexander 2001; Gu et al. 2001); b) Alternatively, Tran (2001, 2003,
also see Yu \& Hwang 2005) 
suggested that at least some of the Sy2s without PBL are
powered by starburst rather than accretion onto a supermassive
black hole, therefore, the BLRs simply do not exist; 
c) More luminous sources tend to have large scaleheight of the scattering
region thus increasing the visibility of PBL (Lumsden \& Alexander 2001);
d) Nicastro, Martocchia \& Matt (2003) have argued that at very low 
accretion rates (and therefore lower luminosities) the clouds of the BLRs 
would cease to exist and the absence of PBL in Sy2s is consistent to their 
low accretion rates.
e) In the case of low luminosity nuclei,
the adjacent bright sources can easily outshine the nuclear flux and
the N$_{\rm H}$ derived from X-ray spectrum may be underestimated (e.g.
Georgantopoulos \& Zezas 2003).
f) The large-scale dusty environment (Panessa \& Bassani 2002) or
complex and variable obscuring material (Matt 2000a; Risaliti 2002)
may to some extent affect the appearance of PBL in Sy2s.
g) Long term large amplitude variation in the nuclei activity could vary
the PBL flux and thus the detectability of PBL (Lumsden et al. 2004).

Meanwhile, it's worth to note that there are also weak evidences showing
different absorption in two types of Sy2s: 
Gu, Maiolino \& Dultzin-Hacyan (2001) found slightly (but not statistically 
conclusive) lower N$_{\rm H}$ for PBL Sy2s, and
Lumsden, Alexander \& Hough (2004) found a considerably higher detection 
rate of scattered broad H$_{\alpha}$ 
in a small sample of Compton-thin Sy2s.
Note one must be cautious while comparing the fraction of PBL Sy2s
between samples since the luminosity might have played a major role.
These evidences suggest that besides the AGN luminosity,
the X-ray absorption column density also plays a role in the visibility of PBL
in Sy2s.
We point out that the role of the absorption (if exists)
could reveal itself in a luminous Sy2 sample where 
the influence of luminosity on the visibility of PBL is weak enough.

In this paper we revisit the issue of whether the nuclear obscuration
in Seyfert 2 galaxies affects the visibility of PBL by focusing on
luminous Sy2s.
The launch of the $Chandra$ X-ray Observatory in 1999 and XMM-$Newton$ in 2000
open a new era of X-ray astronomy. New $Chandra$ and XMM-$Newton$ observations
have significantly enlarged the sample of Sy2s with both spectropolarimetric and
X-ray observations available, and also provide more reliable X-ray measurements
thanks to their much higher spatial resolution and better sensitivity.  
In this paper, we present a large sample of Sy2s, for which both the 
spectropolarimetric and X-ray data are available, to probe the nuclear
obscuration for PBL/NPBL Sy2s. Our sample consists
of 29 PBL Sy2s and 25 NPBL Sy2s. Among them 8 $Chandra$ and 30 XMM-$Newton$
observations are available either from literature or from archive.
Throughout this paper we use the cosmological parameters H$_{0} = 70$ km s$^{-1}$ Mpc$^{-1}$, $\Omega_{m} = 0.27$ and $\Omega_{\lambda} = 0.73$. \\

\section{Sample selection}

We collect all Sy2s with spectropolarimetric observations from 
literature (from 1985 to 2006, see table 1). 
The spectropolarimetric data are 
mainly from several large surveys including: the infrared-selected sample of Heisler et al. (1997), the far infrared flux and luminosity limited sample of Lumsden et al. (2001), the distance limited sample of Moran et al. (2000, 2001), the heterogeneous optical and mid infrared selected sample of Tran et al. (2001), and infrared color selected sample of Young et al. (1996). 
We then exclude NGC 2992, NGC 5506, NGC 5252, NGC 7314, MCG -3-34-64 and 
Mrk 334\footnote{Among the six sources, NGC 5506 and Mrk 334 do not show PBL 
in the polarized spectra and the rest four do. Including these sources in our sample
does not make significant difference to our major results.} from discussion due to their intermediate classification 
(i.e. Sy1.8s, 1.9s) in NED. 
To avoid luminosity selection bias due to the redshift difference when 
comparing properties between the PBL and NPBL Sy2s, we confine our sample 
within z $<$ 0.06. We collect [O {\sc iii}] $\lambda$5007 and X-ray data from literature, and present
spectra analysis of archive $Chandra$/XMM-$Newton$ data for 8 sources in \S3. 
The result from most recent observation
is adopted
when two or more observations exist. 
Note in the table, there are 8 upper limits in the hard X-ray flux
due to X-ray non-detection in the hard band and their X-ray absorption
column densities in literature were estimated either from soft band X-ray 
data or the strength of their X-ray emission relative to optical band. 
We exclude these N$_{\rm H}$
from our following analysis. NGC 4117 is also excluded since its 
[O {\sc iii}] $\lambda$5007 flux is not available from literature.
This leaves a sample composing of 29\footnote{For one galaxy IRAS 04385-0828, the value of N$_{\rm H}$ is unavailable, but we can get the reference to hard X-ray flux from Polletta et al. (1996). When using the "T" ratio for analysis, the number for PBL Sy2s is 30.} Sy2s with PBL, and 25 Sy2s without PBL, for which both the spectropolarimetric and X-ray data are available.  Fig. 1 shows the redshift against the luminosity of  extinction-corrected [O {\sc iii}] $\lambda$5007 emission for PBL and NPBL Sy2s
in the sample. 
As previous studies have shown, we clearly see higher luminosities for PBL Sy2s
(with a confidence level of 99.7\%), indicating the nulear activity plays a 
major role in the visibility of PBL in Sy2s.

The optical and X-ray data are presented in Table 1. 
The table lists, in turn, the name of the galaxy, redshift $z$ as reported 
in NED, the spectropolarimetric properties, the corresponding references, the 
extinction-corrected flux of [O {\sc iii}] $\lambda$5007 emission in units 
of 10$^{-12}$ erg s$^{-1}$ cm$^{-2}$, the luminosity of extinction-corrected [O {\sc iii}] $\lambda$5007 emission in units of erg s$^{-1}$, 
the references for the [O {\sc iii}] $\lambda$5007 emission, the observed rest frame hard X-ray (2-10 keV) flux in units of erg s$^{-1}$, the X-ray absorption column density (N$_{\rm H}$) in units of cm$^{-2}$, the equivalent width (EW) of the fluorescence iron line in units of eV, the references for the X-ray properties. The luminosity of extinction-corrected [O {\sc iii}] $\lambda$5007 emission is given as $\rm L_{\rm [O~III]} = 4 \pi D^{2}F_{[O~III]}^{cor}$, 
where $\rm F_{[O~III]}^{cor}$ is the extinction-corrected flux of 
[O {\sc iii}] $\lambda$5007 emission derived from the relation 
(Bassani et al. 1999) \[ F_{\rm
[O~III]}^{\rm cor} = F_{\rm [O~III]}^{\rm obs} \rm
(\frac{(H_\alpha/H_\beta)_{obs}}{(H_\alpha/H_\beta)_{0}})^{2.94}
\]
We assume an intrinsic Balmer decrement $\rm (H_\alpha/H_\beta)_{0} = 3.0$.

\section{X-ray spectral analysis}

In this section, we report the results of X-ray spectral fitting to
archive $Chandra$ and XMM-$Newton$ spectra of 8 Sy2s in the sample.
The data were reduced using CIAO 3.2.2 and XMMSAS
6.5.0 respectively. The size of each source on the detector was
estimated in order to determine appropriate source extraction regions,
typically $\sim 2"$ radius ($Chandra$) or $\sim 30"$ (XMM-$Newton$)
for on-axis point sources. The background spectra were extracted from source-free annulus around the
source. The spectra of each galaxy were binned to a minimum of 1 counts
per bin and we adopt the  C-statistic (Cash 1979) for minimization. 
Spectral fits were performed using XSPEC version 11.2 in the 0.5 -- 8 keV band.
All the quoted errors are $90\%$ confidence range for one parameter
of interest.

Each spectrum was initially fitted with a simple model consisting of 
a powerlaw plus Galactic and intrinsic neutral absorption. In many cases this simple parameterization is not sufficient to model the 
whole 0.5 -- 8 keV spectrum. Residuals often show a soft excess on 
top of the powerlaw. The soft excess is fitted here as
a scattered powerlaw component 
(with the same powerlaw slope but no intrinsic absorption). 
The possible presence of a narrow emission line centered at 6.4 keV originating from neutral iron has also been checked, and modeled with 
a single Gaussian line. 

We note that in Compton-thick sources with N$_{\sc H}$ $>$ 10$^{24}$ cm$^{-2}$, the
transmitted component is heavily suppressed below 10 keV and the spectrum
observed in the 2-10 keV band might be dominated by the reflection component
(Matt et al. 2000b). In this paper, NGC 34, NGC 3982, Mrk 573, Mrk 1066 are classified as Compton-thick
based on their large Fe K$\alpha$ EW ($>$ 1keV except for NGC 34, see notes in
Appendix) and 
small $\rm F_{2-10~keV}/F_{[O~III]}$
ratios ($<$0.1) (Maiolino et al. 1998, Bassani et al. 1999, 
Guainazzi et al. 2005b). For Compton-thick sources, we use the reflection model ($pexrav$ model in
XSPEC; Magdziarz \& Zdziarski 1995) for spectrum fitting.

Given the purpose of this work (to obtain a proper description of the spectra in 
terms of absorption, 2-10 keV flux, and Fe K line intensity), these simple 
parameterizations yield adequate fits to all the spectra presented here. 
The best-fit spectral parameters are listed in Table 2 and notes on 
individual objects are given in Appendix.
 
\section{different obscuration in two types of Sy2s}

In Fig. 2 (b), we plot the luminosity of extinction-corrected [O {\sc iii}] $\lambda$5007 emission versus N$_{\rm H}$. The separation is 
apparent for two types of Sy2s. The diagram can be roughly divided into 
three regions with the boundaries at N$_{\rm H}$ = 10$^{23.8}$ cm$^{-2}$ and 
L$_{\rm [O~III]}$ = 10$^{41}$ erg s$^{-1}$. For the luminous Sy2s 
(with L$_{\rm [O~III]}>10^{41}$erg s$^{-1}$), we can clearly see that most 
of the Sy2s with N$_{\rm H}<$10$^{23.8}$ cm$^{-2}$ show PBL (86\%, 12 out 14), 
while the fraction is much smaller for sources with heavier obscuration
(54\%, 15 out 28). 
For the Sy2s with lower [O {\sc iii}] luminosity ($<$ 10$^{41}$ erg s$^{-1}$), 
only a small fraction show PBL (17\%, 2 out of 12), and due to the 
limited number of sources, we are not able to tell if the fraction of
PBL sources depends on the X-ray absorption at lower luminosity.
                                                                              
In Fig. 3 (left panel) we plot the N$_{\rm H}$ distributions for all Sy2s 
with/out PBL. Since there are 11 censored data (lower limits) among PBL 
Sy2s and 10 among NPBL Sy2s, we use the survival analysis methods 
ASURV (Feigelson \& Nelson 1985) for statistical analysis. 
We find little difference (with a confidence level of 
66.5$\%$ of the difference, see table 3) in N$_{\rm H}$ between PBL/NPBL Sy2s 
and the mean values of log N$_{\rm H}$ (in units of cm$^{-2}$) are 
23.755$\pm0.19$ and $23.852\pm0.274$, respectively (for NGC 4501 and NGC 
7590 we adopt the N$_{\rm H}$ upper limits as the measured values since 
ASURV could not deal with the case which 
contains both upper and lower limits). However, if we only consider the 
luminous Sy2s with L$_{\rm [O~III]}$ $>$ 
10$^{41}$ erg s$^{-1}$, K-S test shows that the probability for two samples 
to be extracted from the same parent population is about 7.7$\%$, and the 
mean values of log N$_{\rm H}$ are $23.739\pm0.212$ and $24.428\pm0.192$,
respectively (Fig. 3, right panel). The results suggested that for luminous 
Sy2s in our sample, sources without PBL show larger obscuration than those 
with PBL with a confidence level of 92.3\%. 

To further examine if obscuration plays a role in the detection/visibility of
PBL in Sy2s, we explore other potential measures of obscuration. By studying 
a large sample of Sy2s, Bassani et al. (1999) found that 
the "T" ratio $\rm F_{2-10~keV}/F_{[O~III]}$ is a good indicator of nuclear 
obscuration. In particular, it is anticorrelated with both the column density 
N$_{\rm H}$ and 
the Fe K$\alpha$ line EW, and that these quantities can be used as probes 
of the obscuration to the center of the AGN. In Fig. 2 (c, d) we plot the 
Fe K$\alpha$ line EW and "T" ratio versus the luminosity of 
extinction-corrected [O {\sc iii}] $\lambda$5007 emission. Similar patterns 
as seen in Fig. 2 (b) are also obvious that for luminous Sy2s, the NPBL 
sources tend to be more obscured. Fig. 4 shows the "T" ratio distributions 
for all Sy2s (left) and for luminous objects only (right). A K-S test shows 
that the possibility for these two samples to be extracted from the same parent 
population is about 25\%. The mean value of the log T are -0.087$\pm$0.145 
and -0.342$\pm$0.217 for 
Sy2s with and without PBL respectively. 
Similarly the confidence level for the difference is much higher (at level of 99.1\%) for 
luminous Sy2s only. Turning to the  Fe K$\alpha$ line EW, K-S tests 
also confirms that for luminous Sy2s the difference between the two sample is
present (at 95.3\% level) with available data. The mean values of log EW(Fe) 
are 2.626$\pm$0.107 and 2.999$\pm$0.066 respectively.
After examining three independent indicators for obscuration, we 
conclude that for luminous Sy2s, sources without PBL have higher obscuration 
than those with PBL, confirming the suggestion that the obscuration does play 
an important role in the detectability/visibility of PBL.  
The results from K-S tests and average values for N$_{\rm H}$, "T" ratio and FeK line
EW are summarized in table 3.

\section{Discussion}
It's clear that, as many previous studies have shown, PBL Sy2s have higher
luminosities than NPBL Sy2s (see Fig. 1 \& 2), indicating the primary 
determinant of PBL visibility is the nuclei luminosity.
In this paper, by focusing on luminous Sy2s, we find that the nuclei obscuration
also plays an important role in the visibility of PBL. 
For Sy2s with L$_{\rm [O~III]}$ $>$
10$^{41}$ erg s$^{-1}$ in our sample, we find that NPBL Sy2s have higher
X-ray column density than PBL Sy2s at a significant level
of 92.3\%.
While using the "T" ratio or the Fe K line EW as 
indicator of nuclear obscuration, the confidence level of the difference
in obscuration gets even higher (99.1\% and 95.3\% respectively). Our results are consistent 
with Lumsden et al. (2004)
who reported higher detection rate of PBL in Compton-thin Sy2s, but with
much higher confidence level. Consistent with previous studies,
most (83\%) of the less luminous Sy2s (L$_{\rm [O~III]}$ $<$ 10$^{41}$ erg 
s$^{-1}$) do not show PBL, the nature of which is still unclear
(see \S1) and is beyond the scope of this paper. 
We also demonstrate that since most of the less luminous Sy2s do not show PBL 
independent of absorption, adding them to the sample
weakens the difference in obscuration found in the luminous sample.
This explains why previous studies, which did not exclude less luminous sources,
found no difference in absorption.

\subsection{Selection effect?}
It is worth stressing whether the difference in the nuclear obscuration between 
two types of Sy2s could be due to possible observational and sample selection 
bias.
We note our sample is an amalgamation of different observations with diverse 
quality of spectropolarimetric data. We first examine whether the 
non-detections of PBL in the sample are due to the weakness/lack of PBL or
due to the limited sensitivities of the spectropolarimetric data.
Lumsden et al. (2001) showed that the S/N in their sample is sufficient 
for all but 2 of the NPBL S2s and attributed
the nondetections to significantly weaker scattered flux. 
Tran (2003) pointed out that the 
distributions of [O {\sc iii}] flux, which is a good indicator of the 
strength of the Seyfert nucleus, are virtually the same between two types of
Sy2s in his sample,
suggesting that the nondetections are not likely due to the detection limit
of the survey. The sensitivity of Moran's sample (2001) is found to be 
even better than that of the other samples mentioned above (Gu \& Huang 2002).
In Fig. 5, we plot the "T" ratio vs. the extinction corrected [O {\sc iii}] flux
for our composite sample. We can clearly see that for luminous sources (with L$_{\rm [O~III]}$ $>$ 10$^{41}$ erg s$^{-1}$) in our sample, there
is no difference in the [O {\sc iii}] flux distributions between PBL and 
NPBL Sy2s. For comparison, sources with lower luminosities are also plotted,
most of which are much weaker in the [O {\sc iii}] flux.
We conclude that most of the nondetections for our luminous sources are 
likely due to the weakness or lack of PBL but not
due to the limited sensitivities of the spectropolarimetric data.

We then verify if our results are biased by combining samples with 
different selection criteria and survey depths into one single sample, i.e.,
if some of samples tend to select more obscured sources but with poor 
spectropolarimetric data, and/or some others tend to select less obscured 
sources but with better spectropolarimetric data.
By plotting in Fig. 6 the "T" ratio vs. [O {\sc iii}] luminosity for sources in each 
subsample, we can see this is not the case for our composite sample.
We find that each subsample spans a similar obscuration range with
that of the composite sample, and the difference in 
the obscuration between PBL and NPBL Sy2s are also visible in most of the 
subsamples. 

We made an additional test to check whether the distributions of $z$ and 
[O {\sc iii}] luminosity for our luminous sample are different (see Fig. 
1 and 2 (a)). 
Using ASURV, we get average values of $\langle z \rangle$ = 0.02$\pm$ 0.003 and $\langle \rm log L_{[O~III]}\rangle$ = 42.198 $\pm$ 0.121 for PBL Sy2s whereas $\langle z \rangle$ = 
0.017$\pm$ 0.004 and $\langle \rm log L_{[O~III]}\rangle$ = 41.991 $\pm$ 0.157 for NPBL 
Sy2s. The distributions of $z$ and [O {\sc iii}] luminosity for luminous sources are similar 
at levels of $\rm p_{null}$ = 29.4\% and $\rm p_{null}$ = 30.6\%, respectively. 
The similarity indicates that the difference in absorption could not be 
biased by different redshift/luminosity which both affect the visibility of PBL.
Also, the dilution effect, which is dependent of the redshift and luminosity, 
might bias the visibility of PBL for less luminous Sy2s, but itself alone can 
not explain the difference in the absorption between PBL and NPBL Sy2s.
Actually we note that the dilution effect to the visibility of PBL is much
weaker for our luminous sources. This can be seen from the
fact that PBL Sy2s can be detected in most of luminous
Sy2s with smaller obscuration (N$_{\rm H}$ $<$ 10$^{23.8}$ cm$^{-2}$ or "T" ratio
$>$ 10$^{-0.7}$). The dilution effect in X-ray (to the measurement of
N$_H$) is also much weaker for luminous Sy2s.
We conclude that there is no observational bias which can produce the 
difference in absorption between two types of Sy2s in our sample,
and a physical link between the visibility of PBL and nuclear obscuration is 
required. 

\subsection{physical explanation to the difference in absorption}
  
The results presented here for luminous Sy2s can be interpreted within the 
context of the 
unified model for Seyfert galaxies, in agreement with the torus geometry 
portraited by Heisler et al. (1997): the main electron scattering is 
confined to a conical region that is close to the 
thickness of the torus. More inclined Sy2s could have the broad
line scattering screen also obscured thus make PBL weaker or nondetectable. 
According to the unified model, more inclined sources expect heavier
obscuration, thus explain the difference in obscuration between PBL and
NPBL Sy2s. 

Note the high detection rate (86\%) of PBL in luminous Sy2s with 
N$_H$ $<$ 10$^{23.8}$ cm$^{-2}$ suggests that in most sources the 
Compton-thin X-ray obscuring material can not have much larger scale than the 
scattering screen, supporting the torus scheme of the unified model.
Furthermore, while
extended obscuration from the host galaxy might explain the absence
of PBL in some sources (such as NGC 5506, see Lumsden et al. 2004), it could 
not be the major cause, otherwise we should have seen a large number of NPBL 
Sy2s among luminous Sy2s with N$_H$ $<$ 10$^{23.8}$ cm$^{-2}$.
It's interesting to note that NGC 5506 is one of two intermediate Seyferts 
without PBL detected (the other one is Mrk 334, see \S2), both with the BLR 
visible in near infrared. This suggests extended obscuration from the host 
galaxy is also a plausible cause in the absence of PBL in Mrk 334.

We also note that while using the "T" ratio as indicator of nuclear obscuration,
the difference between two types of Sy2s becomes more significant (see Fig. 2 (d)).
We point out that this is mainly because for Compton thick sources, we can only
give lower limits of $\sim$ 10$^{24}$ cm$^{-2}$ for N$_{\rm H}$, but the "T" 
ratio is a
continuous variable as long as they are detected in the X-ray band.
Fig. 7 shows the observed 2-10 keV X-ray flux against the extinction-corrected 
[O {\sc iii}] flux for Compton-thick Sy2s (with a lower limit of 
10$^{24}$ cm$^{-2}$ of N$_{\rm H}$, 11 PBL and 10 NPBL Sy2s). 
Interestingly, we find although the lower limits of N$_{\rm H}$ are the same for
two types of Compton thick Sy2s, NPBL Sy2s in the figure tend to be weaker in X-ray (with a
confidence level of 97.6\%).
We note that two types of Compton-thick Sy2s have similar large FeK line EW 
($>$ 1 keV)
suggesting
the X-ray spectra in both types are reflection dominated. In this case,
smaller "T" ratio in NPBL Sy2s can also be explained by higher inclination:
sources viewed at higher inclination could have a large fraction of inner 
surface of the torus, where the reflection component is produced, blocked 
from our line of sight, thus expect weaker X-ray emission.

\acknowledgments

This work was supported by Chinese NSF through NSF10473009/NSF10533050, and 
the CAS "Bai Ren" project at University of Science and Technology of China. 
The work in this paper has made use of the NASA's Astrophysics Data System 
Abstract Service and the NASA/IPAC Extragalactic Database (NED), which is 
operated by the Jet Propulsion Laboratory, California Institute of Technology, 
under contract with NASA. This study has also made use of the HEASARC on-line 
data archive services, supported by NASA GSFC.

\appendix

\section{Notes on Individual Objects}

In this section, we present brief discussions for the X-ray data of 3 sources in the sample. 
All the errors quoted are at the 90\% level of confidence. 

NGC 34: we model the $XMM-Newton$ 0.5 -- 8.0 keV spectrum of this source with 
$pexrav$ model (C/dof = 362/448) in terms of its low "T" ratio of 0.03 ($<$0.1).
However, the Fe K line is marginally detected with a upper limit EW of 321 eV. 
We then fit the spectrum with an absorbed powerlaw model which gives $\Gamma$ 
= 1.68$^{+0.11}_{-0.17}$ but no intrinsic absorption, plus Fe K line with EW 
= 386 ($<1047$) eV. The model yields a worse fit with C/dof = 389/448. We note 
the steep spectrum slope may attribute to the host galaxy thermal emission in 
the soft X-ray band. From the lower "T" ratio of 0.03 ($<$0.1) and better C 
statistic of $pexrav$ model, we consider the galaxy as Compton-thick and give a lower limit of 10$^{24}$ cm $^{-2}$ to N$_{\rm H}$.

NGC 5728: The $Chandra$ 0.5 -- 8 keV spectrum is parameterized here with an 
absorbed powerlaw ($\Gamma$ = 2.73$\pm$0.17, N$_{\rm H}$ = 
7.8$^{+1.5}_{-1.4}$ $\times$ 10$^{23}$ cm$^2$) plus a 0.4\% scattered component.
The Fe K line is detected at 6.4 keV with EW = 1100$^{+320}_{-270}$ eV. For the 
lower "T" ratio and large Fe line EW, we then fit the spectrum of this galaxy with 
$pexrav$ model plus Gaussian line. However, the fitting is unacceptable (C/dof 
= 710/300). So we do not regard it as a Compton-thick one in this paper. We 
note considering it as Compton-thick does not affect our results presented here. 

NGC 6552: We fit the $XMM-Newton$ 0.5 -- 8 keV spectrum of this source by an 
absorbed powerlaw ($\Gamma$ = 2.8$^{+0.37}_{-0.13}$) with a 0.75\% scattered 
component. The best fit (C/dof = 121/166) gives N$_{\rm H}$ = 
7.1$^{+4.0}_{-1.0}$ $\times$ 10$^{23}$ cm$^2$. The Fe K line is detected at 
6.4 keV with EW = 1408$^{+668}_{-883}$ eV. However, the $pexrav$ model can 
also give an acceptable fitting of the spectrum with  $\Gamma$ = 
2.86$^{+0.34}_{-0.47}$ and Fe K line EW = 4990$^{+3910}_{-2390}$ 
(C/dof = 128/166). The fitted 2-10 keV flux is 2.32$\times$10$^{-13}$ 
erg s$^{-1}$ cm$^{-2}$. We adopt the absorbed powerlaw model for the spectrum 
fitting in the paper in terms of the better C statistic. We note that the 
consideration of it as Compton thick will not affect our results presented 
in the paper.

\begin{figure}
\epsscale{.50}
\plotone{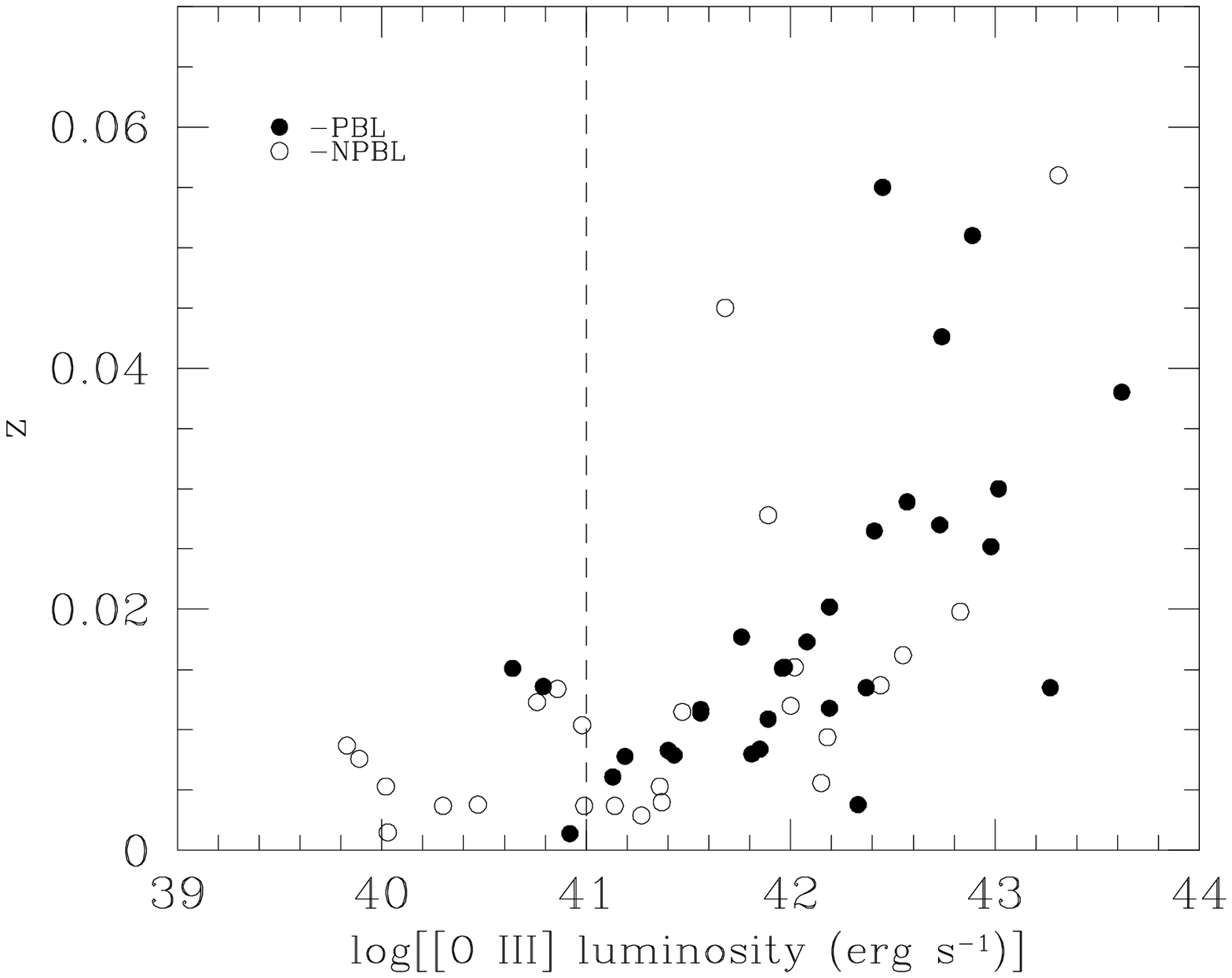} 
\caption{Redshift vs. [O {\sc iii}] $\lambda$5007 luminosity for two types of Sy2s in the sample. solid circles stand for PBL Sy2s and open circles for NPBL Sy2s.
\label{fig1}} 
\end{figure}

\clearpage

\begin{figure}
\epsscale{1.0}
\plottwo{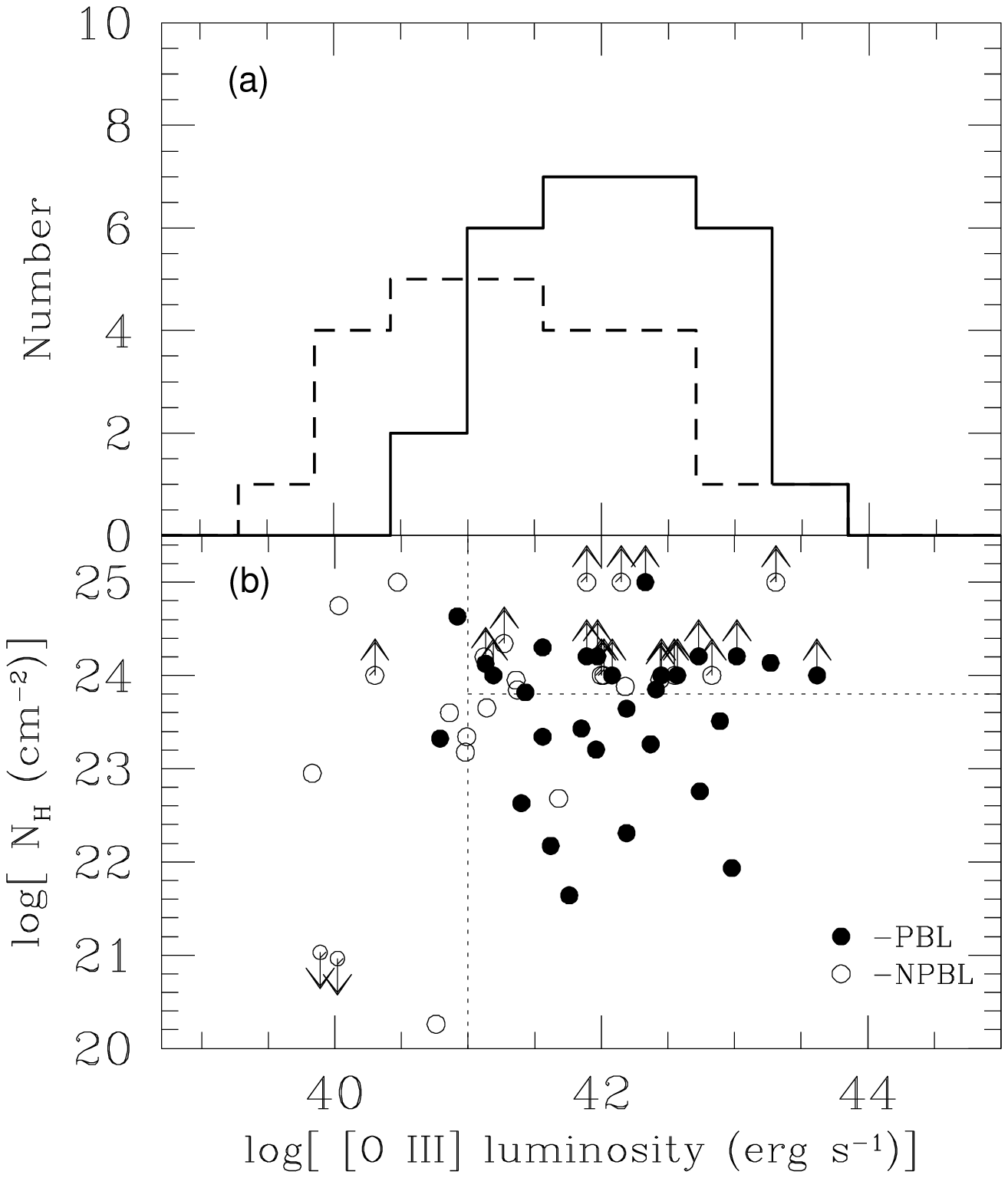}{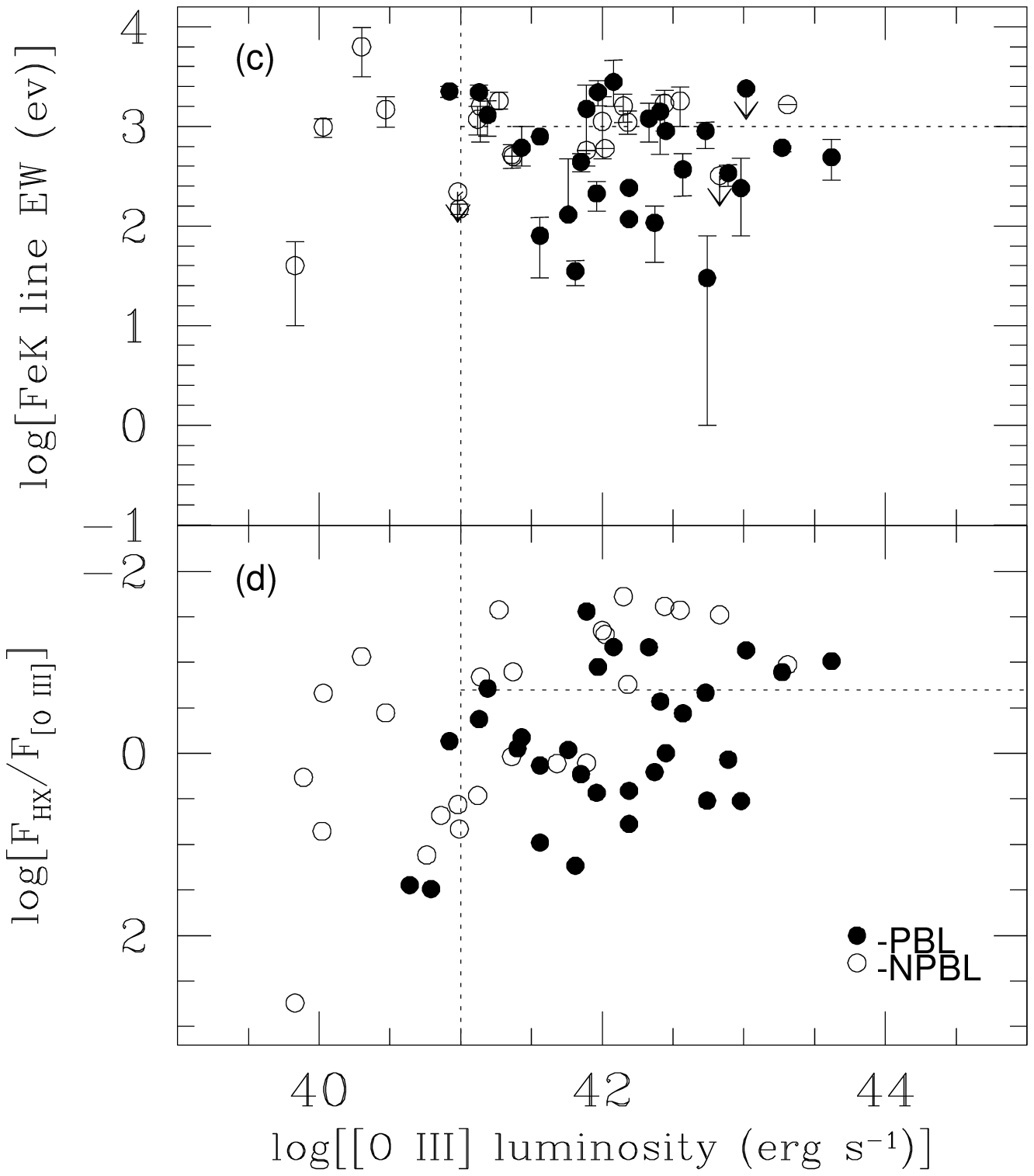}
\caption{ The left bottom panel plots the absorption column density N$_{\rm H}$ 
versus the [O {\sc iii}] $\lambda$5007 luminosity for PBL and NPBL Sy2s.
The distribution  of the [O {\sc iii}] $\lambda$5007 luminosity for two types 
of Sy2s is presented in the left top panel (solid line for PBL Sy2s and dashed 
line for NPBL Sy2s). The right panel is the plot of FeK line EW (top) and 
"T" ratio (bottom) versus the [O {\sc iii}] $\lambda$5007 luminosity.
.\label{fig2}}
\end{figure}

\clearpage
\begin{figure}
\epsscale{1.0}
\plottwo{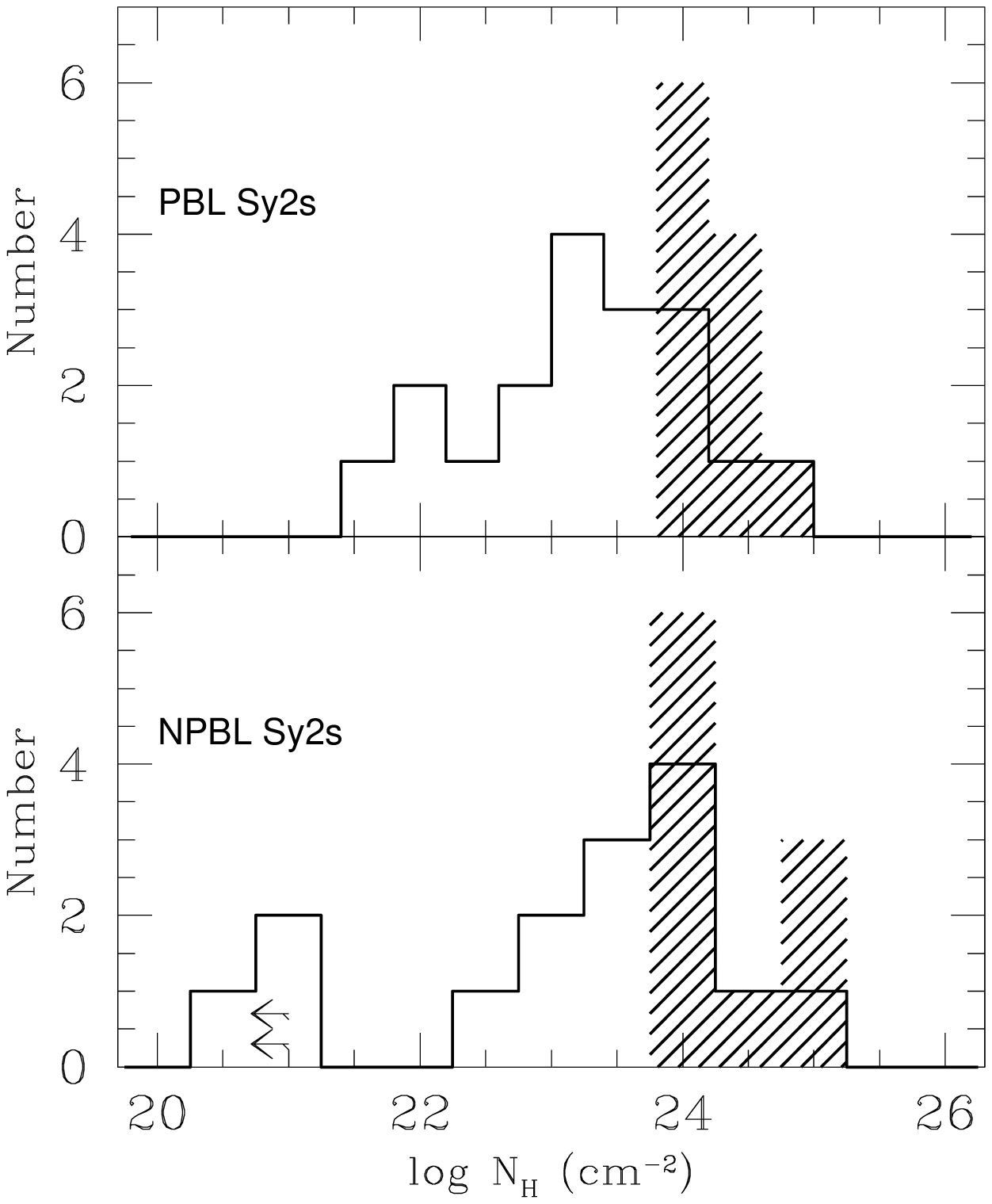}{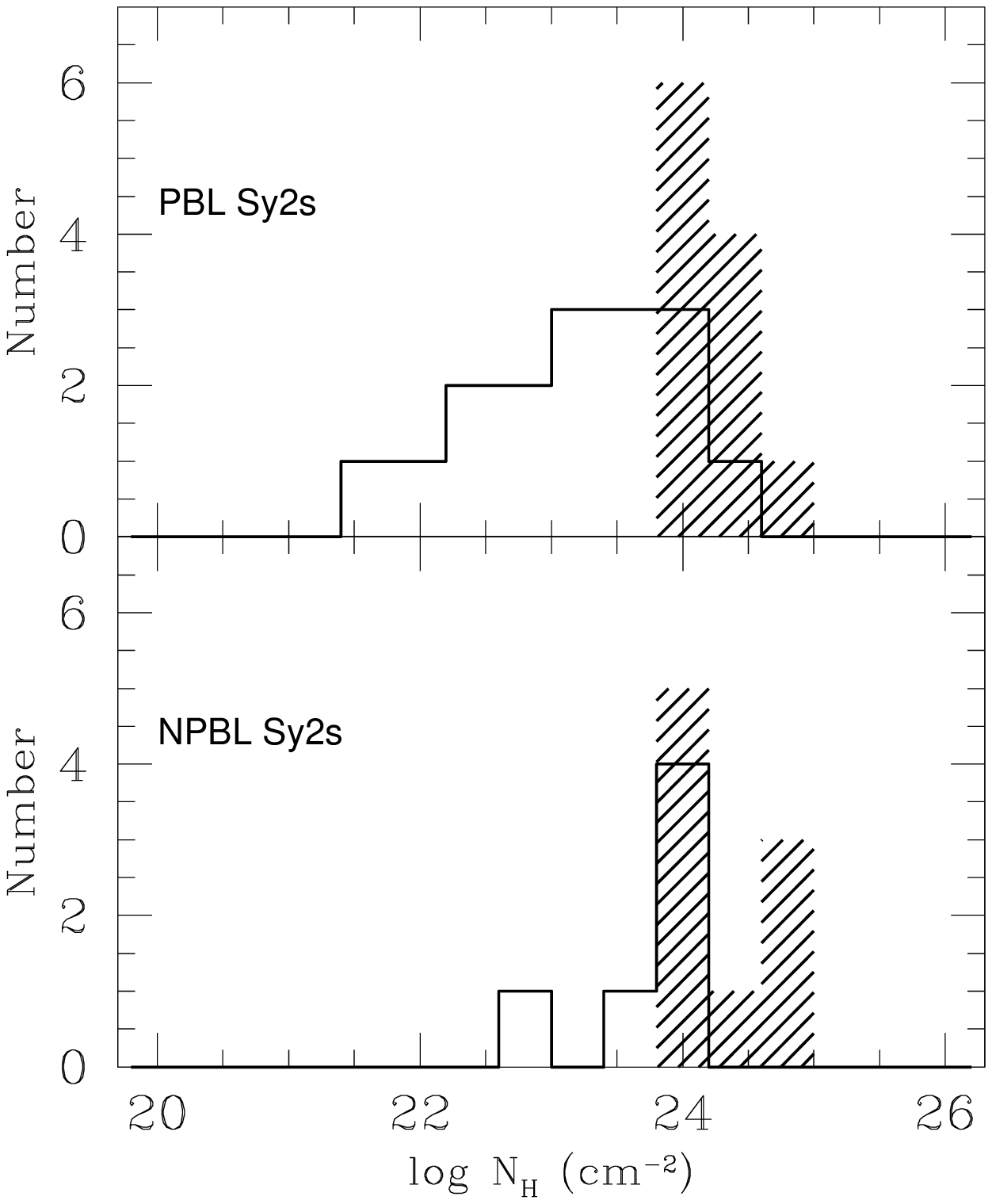}
\caption{N$_{\rm H}$ distributions for all the Sy2s in our sample (left) and 
luminous sources with log L$_{\rm [O~III]}$ $>41$ erg s$^{-1}$ (right). Shaded 
areas denote lower limits. The arrows denote the upper limits of N$_{\rm H}$.
}
\end{figure}

\clearpage
\begin{figure}
\epsscale{1.0}
\plottwo{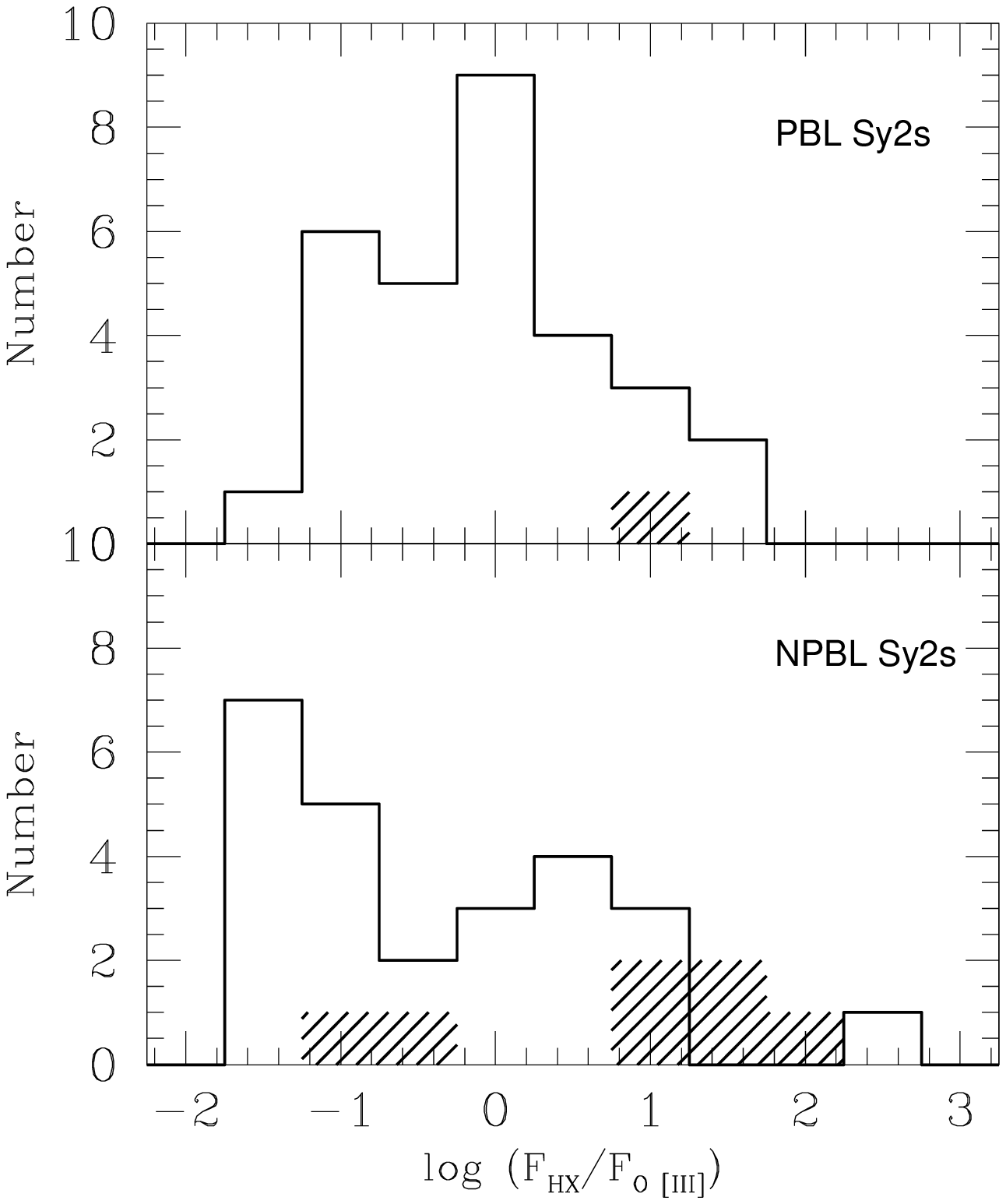}{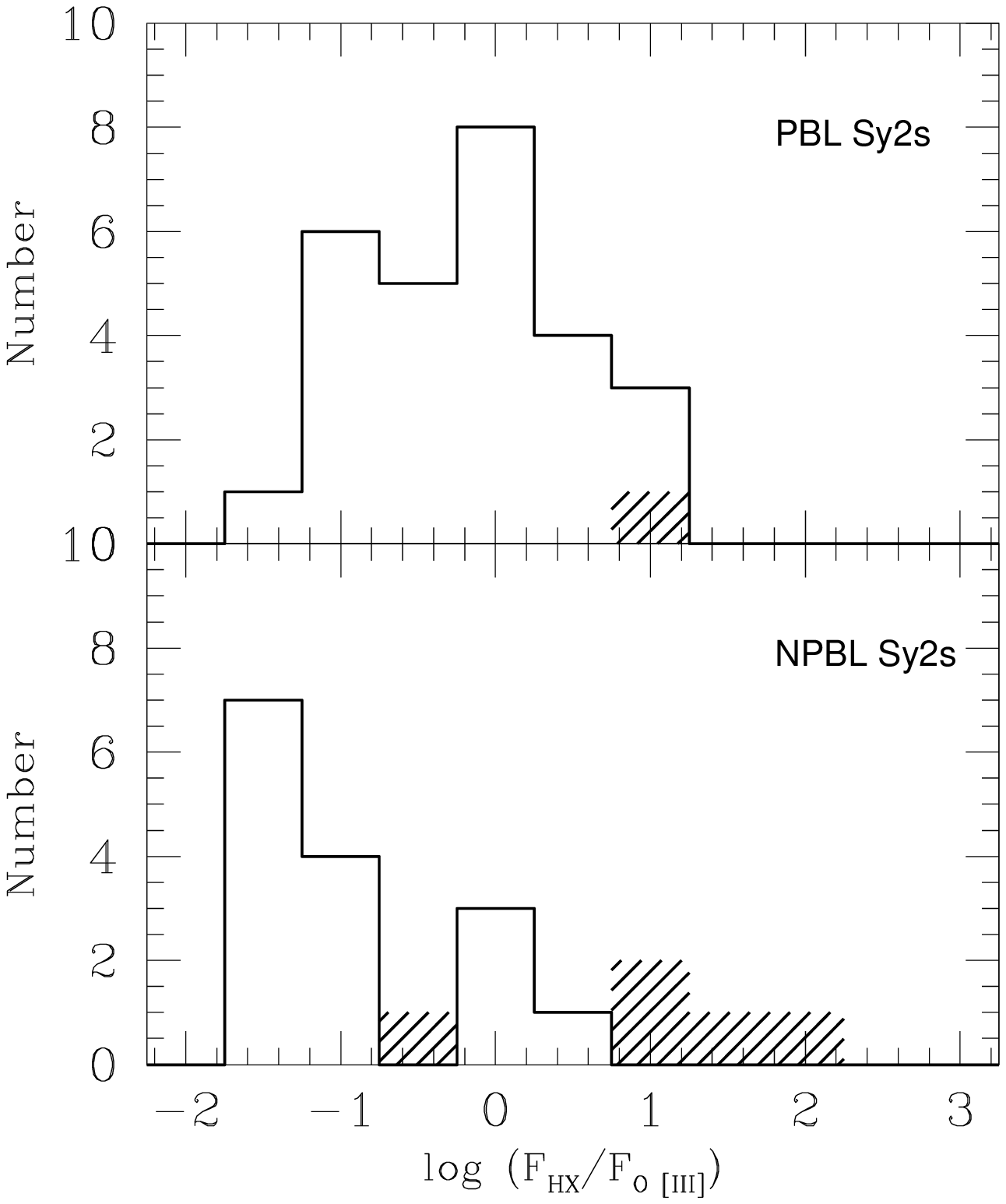}
\caption{Distributions of the "T" ratio for all Sy2s (left) and for luminous 
Sy2s only (right) in our sample. Shaded areas denote the 8 sources with only
hard X-ray upper limits. \label{fig3}}
\end{figure}

\clearpage
\begin{figure}
\epsscale{0.5}
\plotone{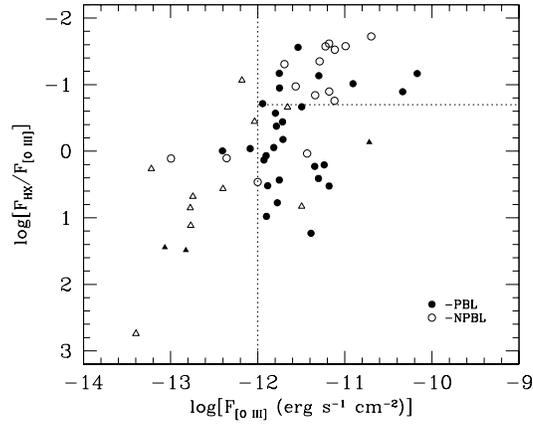}
\caption{The plot of "T" ratio vs. extinction-corrected [O {\sc iii}] flux.
Sources with luminosity L$_{\rm [O~III]}<$10$^{41}$ erg s$^{-1}$ are plotted 
as triangles. \label{fig5}}
\end{figure}

\clearpage
\begin{figure}
\epsscale{0.5}
\plotone{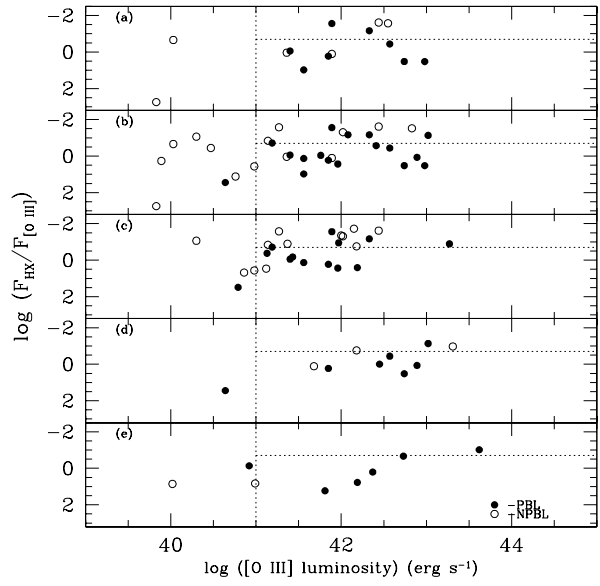}
\caption{ The plot of "T" ratio vs. [O {\sc iii}] $\lambda$5007 luminosity for different spectropolarimetric subsamples. (a): Lumsden et al. (2001)'s sample, (b): Tran (2001)'s sample, (c): Moran et al. (2000)'s sample, (d): Young et al. (1996)'s sample, (e): other surveys.  \label{fig6}} 
\end{figure}

\clearpage
\begin{figure}
\epsscale{0.5}
\plotone{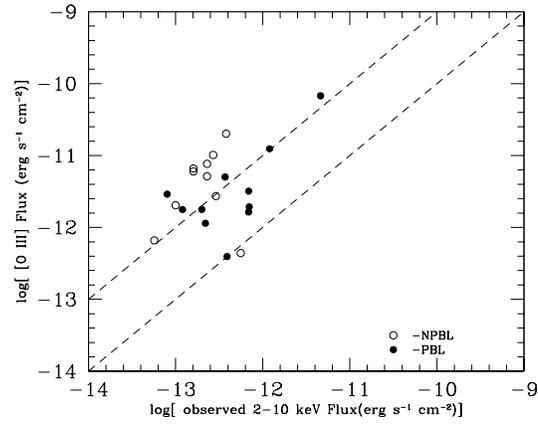}
\caption{The observed [O {\sc iii}]  flux (extinction corrected) against 
X-ray (2 -- 10 keV) flux for Compton-thick Sy2s (with N$_{\rm H}$ $>$ 10$^{24}$ 
cm$^{-2}$). The dashed lines represent F$_{\rm [O~III]}$ = 10 F$_{2-10keV}$
(upper) and F$_{\rm [O~III]}$ = F$_{2-10keV}$ (lower).
\label{fig7}}

\end{figure}

\clearpage

\begin{deluxetable}{lcccccccccc}
\tabletypesize{\scriptsize}
\rotate
\tablecaption{\scshape Optical and Hard X-Ray Data for Seyfert 2 Galaxies with/out PBL}
\tablewidth{0pt}
\tablehead{
\colhead{Name} & \colhead{z} & \colhead{PBL?}  & \colhead{References} & \colhead{F$_{\lambda5007}$} & \colhead{L$_{\rm [O~III]}$} & \colhead{References} & \colhead{F$_{\rm 2-10~keV}$} & \colhead{log$_{\rm 10}$ N$_{\rm H}$} & \colhead{EW(Fe)} & \colhead{References} \\
\hspace*{10.mm}(1) & (2) & (3) &(4) & (5) & (6) & (7) & (8) & (9) & (10) & (11) }
\startdata
Lumsden et al. (2001)'s sample: \\ 
\\
Mrk 334     & 0.022 & n  & 66  & 2.0  & 42.34 & 22 & $<$13 & 20.643 & $\dots$ &  36 \\
IRAS 00198-7926 & 0.0728 & n  & 2  & 0.36 & 42.67 & 16 & $<$0.1 & $>$24 & $\dots$ & 62 \\
NGC 1068        & 0.0038 & y & 3L & 67.8 & 42.33 & 22 & 4.62 & $>$25 & 1200$\pm$500 & 5  \\
NGC 1143        & 0.0291 & n & 2  & 0.48 & 41.97 & 2  & $\dots$   & $\dots$   &$\dots$& $\dots$  \\ 
IRAS 04259-0440 & 0.0155 & n & 2  & 1.3  & 41.85 & 2  & $\dots$   & $\dots$    &$\dots$ & $\dots$ \\ 
IRAS 05189-2524 & 0.0426 & y  & 1A & 1.3  & 42.74 & 2  & 4.3  & 22.756 & 30$^{+50}_{-30}$ & 2  \\ 
NGC 4388        & 0.0084 & y & 1A & 4.51 & 41.85 & 22 & 7.62 & 23.43 & 440$_{-90}^{+90}$ & 5 \\
IC 3639         & 0.0109 & y & 2 & 2.9  & 41.89 & 27 & 0.08 & $>$24.204 & 1500$_{-1100}^{+1100}$ & 7 \\
MCG -3-34-64    & 0.0165 & y & 1A & 4.0 & 42.39 & 2  & 4.0 & 23.614 &  356$_{-143}^{+186}$ & 67 \\
NGC 5135        & 0.0137 & n & 2,33A & 6.61  & 42.44 & 27 & 0.16 & $>$23.954 & 1700$_{-800}^{+600}$ & 8 \\ 
NGC 5194        & 0.0015 & n & 2,15L  & 2.2 & 40.03 & 31 & 0.48 & 24.748 & 986 $_{-210}^{+210}$ & 5 \\
NGC 5256        & 0.0278 & n & 2,15L & 0.44 & 41.89 & 22 & 0.56 & $>$25 & 575 &  62 \\ 
Mrk 1361        & 0.0226 & n & 2  & 1.8  & 42.32 & 2 & $\dots$    & $\dots$    & $\dots$ & $\dots$ \\
NGC 5929$^*$  & 0.0083 & y & 42K & 1.53 & 41.40 & 2 & 1.35 & 22.629 & $\dots$ & 9\\
NGC 5995        & 0.0252 & y  & 2 & 6.6  & 42.98 & 2 &  22   & 21.934  & 240$_{-160}^{+240}$ & 2 \\
IRAS 19254-7245$^*$ & 0.0617 & y  & 69E & 1.26 & 43.06 & 68 & 0.23 & $>$24 & $2000\pm600$ &70 \\ 
IC 5063         & 0.0114 & y & 10A & 1.26 & 41.56 & 19 &  12  & 23.342 & 80$_{-50}^{+42}$ & 49 \\
NGC 7130        & 0.0162 & n  & 2   & 6.0  & 42.55 & 27 & 0.16  & $>$24  & 1800$_{-800}^{+700}$ & 12 \\
NGC 7172        & 0.0087 & n  & 2 & 0.04 & 39.83 & 35  & 22 &  22.95 & $40\pm30$ & 13 \\
IC 5298         & 0.0273 & n & 2  & 1.7  & 42.46 & 2  & $\dots$ & $\dots$  & $\dots$ & $\dots$ \\ 
NGC 7582     & 0.0053 & n & 33A,2 & 3.69 & 41.36 & 27 & 4.0 & 23.95 & 521$_{-141}^{+139}$ & 60 \\
NGC 7674        & 0.0289 & y & 1A,20L & 1.93 & 42.57 & 22 & 0.7 & $>$ 24 &  370$_{-170}^{+160}$ & 14 \\ \\

Tran (2001)'s sample: \\ \\
IRAS 00521-7054 & 0.0689 & n & 1A & 0.36 & 42.62 & 1 &  $<$31.8 & $\dots$ & $\dots$  & 39 \\ 
IRAS 01475-0740 & 0.0177 & y  & 15P & 0.82 & 41.76 & 16 & 0.75  & 21.59 & 130($<$344) & 13 \\
IRAS 02581-1136 & 0.0299 & y &  15L & 0.07 & 41.16 & 22 & $\dots$   &  $\dots$   &  $\dots$ & $\dots$ \\
IRAS 04385-0828 & 0.0151 & y & 15LP & 0.086 & 40.64 & 32 & 2.4  & $\dots$ & $\dots$ & 39 \\
IRAS 05189-2524 & 0.0426 & y  & 1A & 1.3  & 42.74 & 2  & 4.3  & 22.756 & 30$^{+50}_{-30}$ & 2  \\
IRAS 15480-0344 & 0.03 & y & 15P,1A & 5.03 & 43.02 & 16 & 0.37 & $>$24.204 & $<$2400 & 7 \\
IRAS 22017+0319 & 0.0611 & y  & 15P,1A & 0.42 & 42.58 & 1 & 3.6  & 22.69 & $380_{-160}^{+180}$ & 18 \\ 
IC 5063         & 0.0114 & y & 10A & 1.26 & 41.56 & 19 &  12  & 23.342 & 80$_{-50}^{+42}$ & 49 \\
MCG -3-34-64    & 0.0165 & y & 1A & 4.0 & 42.39 & 2  & 4.0 & 23.614 &  356$_{-143}^{+186}$ & 67 \\
Mrk 348         & 0.0151 & y & 20L  & 1.77 & 41.96 & 22 & 4.8  & 23.204   & $212_{-72}^{+68}$ & 21 \\
MCG -3-5-87     & 0.0317 & y & 15P   & 0.37 & 41.93 & 32  & $\dots$ & $\dots$  & $\dots$ & $\dots$ \\
Mrk 463E        & 0.051  & y & 20L,1A & 1.25 & 42.89 &  22 & 1.46 & 23.51 & $340_{-90}^{+70}$ & 6 \\ 
NGC 424         & 0.0117 & y & 23C & 1.18 & 41.56 & 24 & 1.6 & 24.301 & 790 & 25 \\
NGC 513     & 0.0195 & y & 26L & 0.16 & 41.14 & 32 & $\dots$  & $\dots$ & $\dots$ & $\dots$  \\
NGC 1068        & 0.0038 & y & 3L & 67.8 & 42.33 & 22 & 4.62 & $>$25 & 1200$\pm$500 & 5  \\ 
NGC 4388        & 0.0084 & y & 1A & 4.51 & 41.85 & 22 & 7.62 & 23.43 & 440$_{-90}^{+90}$ & 5 \\ 
NGC 5506      & 0.0062 & n & 15 & 3.33 & 41.45 & 58 & 58  & 22.46 & 86$_{-10}^{+24}$ & 11 \\
NGC 5995        & 0.0252 & y  & 2 & 6.6  & 42.98 & 2 &  22   & 21.934  & 240$_{-160}^{+240}$ & 2 \\
NGC 6552        & 0.0265 & y & 15P & 1.6  & 42.41 & 32 & 0.43 & 23.85  & 1408$^{+668}_{-883}$ &  13 \\
NGC 7674        & 0.0289 & y & 1A,20L & 1.93 & 42.57 & 22 & 0.7 & $>$ 24 &  370$_{-170}^{+160}$ & 14 \\ 
NGC 7682        & 0.0171 & y & 15P & 0.87 & 41.76 & 28 & $<$13 & $\dots$ & $\dots$    & 39 \\
IC 3639         & 0.0109 & y & 2 & 2.9  & 41.89 & 27 & 0.08 & $>$24.204 & 1500$_{-1100}^{+1100}$ & 7 \\ 
IRAS 00198-7926 & 0.0728 & n  & 2  & 0.36 & 42.67 & 16 & $<$0.1 & $>$24 & $\dots$ &62 \\
IRAS 03362-1642 & 0.0372 & n & 3L  &  0.13 & 41.62 & 16 & $\dots$  &  $\dots$    & $\dots$ & $\dots$ \\
IRAS 19254-7245$^*$ & 0.0617 & y  & 69E & 1.26 & 43.06 & 68 & 0.23 & $>$24 & $2000\pm600$ & 70 \\ 
NGC 5194        & 0.0015 & n & 2,15L  & 2.2 & 40.03 & 31 & 0.48 & 24.748 & 986 $_{-210}^{+210}$ & 5 \\
NGC 5256        & 0.0278 & n & 2,15L & 0.44 & 41.89 & 22 & 0.56 & $>$25 & 575 &  62 \\
Mrk 573$^*$     & 0.0173 & y & 33S & 1.77 & 42.08 & 22 & 0.12  & $>$24  & 2800$^{+1820}_{-1220}$ & 13 \\
NGC 34      & 0.0198 & n & 15P,33A & 7.68 & 42.83 & 34 & 0.23 & $>$24 & $<$321 & 13 \\
NGC 1144     & 0.0289 & n & 15P,33A  & 0.39 & 41.87 & 34,35 & $<$12 & 20.699 & $\dots$ & 36 \\
NGC 1241        & 0.0135 & n & 15P & 0.91 & 41.57 & 32,35 & $\dots$   & $\dots$   & $\dots$ & $\dots$ \\
NGC 1320       & 0.0094 & n & 15L & 0.57 & 41.05 & 37 & $<$8.2   & $\dots$  & $\dots$ & 39 \\
NGC 1386       & 0.0029 & n & 23 & 10.2   & 41.27 & 38 & 0.27 & $>$24.342 & 1800$^{+400}_{-300}$ & 8 \\
NGC 1667       & 0.0152 & n & 15L,23 & 2.03 & 42.02 & 27   & 0.1 & $>$24 & 600 & 14 \\
NGC 3079       & 0.0038 & n & 15L  & 0.92 & 40.47 & 31  & 0.33 & 25  & 1480$_{-500}^{+500}$ & 5 \\
NGC 3362       & 0.0276 & n & 15L & 0.13 & 41.36 & 39  & $<$ 12.6 & $\dots$  & $\dots$ & 39 \\ 
NGC 3660     & 0.0123 & n & 15L & 0.17 & 40.76 & 27,61 & 2.22 & 20.26 & $\dots$ & 9 \\
NGC 3982       & 0.0037 & n & 15L,23 & 0.66 & 40.3 & 15,31  & 0.057  & $>$ 24 & 6310$^{+3500}_{-3170}$ & 13 \\
NGC 4501       & 0.0076 & n & 15L  & 0.06 & 39.89 & 31 &  0.11   & $<$ 21.03 & $\dots$ &  5 \\
NGC 4941       & 0.0037 & n & 23  & 4.57 & 41.14 & 38 & 0.66     & 23.653  & 1600$_{-900}^{+700}$ & 40 \\ 
NGC 5135        & 0.0137 & n & 2,33A & 6.61  & 42.44 & 27 & 0.16 & $>$23.954 & 1700$_{-800}^{+600}$ & 8 \\
NGC 5283       & 0.0104 & n &15L,23 & 0.4  & 40.98 & 22 & 1.46  & 23.176  &  $<$220 & 7 \\
NGC 5347$^*$       & 0.0078 & y & 42K & 1.14 & 41.19 & 15 & 0.22  & $>$24 & 1300$\pm$500 & 63 \\
NGC 5695       & 0.014  & n & 15L,23  & 0.081 & 40.55 & 22 & $<$0.01 & $\dots$  & $\dots$ & 39 \\ 
NGC 5929$^*$  & 0.0083 & y & 42K & 1.53 & 41.40 & 2 & 1.35 & 22.629 & $\dots$ & 9\\ 
NGC 6890       & 0.0081 & n & 23  & 0.5  & 40.86 & 43 & $\dots$  & $\dots$     & $\dots$ & $\dots$ \\
NGC 7172        & 0.0087 & n  & 2 & 0.04 & 39.83 & 35  & 22 &  22.95 & $40\pm30$ & 13 \\
NGC 7582     & 0.0053 & n & 33A,2 & 3.69 & 41.36 & 27 & 4.0 & 23.95 & 521$_{-141}^{+139}$ & 60 \\
UGC 6100       & 0.0295 & n & 15L & 0.96 & 42.28 & 28 &  $<$11.4  & $\dots$ & $\dots$ & 39\\
\\
Moran et al. (2000)'s sample: \\ \\
IC 3639         & 0.0109 & y & 2 & 2.9  & 41.89 & 27 & 0.08 & $>$24.204 & 1500$_{-1100}^{+1100}$ & 7 \\ 
ESO 428-G014   & 0.0056 & n & 23 & 20.1 & 42.15 & 44 & 0.38 & $>$25 & 1600$\pm$500 & 63 \\
MCG +1-27-020   & 0.0117 & n & 23 & $\dots$ & $\dots$  & $\dots$    & $\dots$    & $\dots$    & $\dots$ & $\dots$ \\ 
Mrk 3        & 0.0135 & y & 20L & 46.1 & 43.27 & 43 & 5.9 & 24.134 & 610$_{-50}^{+30}$ & 45 \\
Mrk 1066       & 0.012  & n & 20L &  5.14 & 42 & 43 & 0.23 & $>$24 &  1120$^{+850}_{-650}$ & 13 \\
Mrk 348         & 0.0151 & y & 20L  & 1.77 & 41.96 & 22 & 4.8  & 23.204   & $212_{-72}^{+68}$ & 21 \\
NGC 424         & 0.0117 & y & 23C & 1.18 & 41.56 & 24 & 1.6 & 24.301 & 790 & 25 \\
NGC 591        & 0.0152 & y & 23K  & 1.78 & 41.97 & 43 & 0.2 & $>$24.204 & 2200$^{+700}_{-600}$ & 7 \\
NGC 788        & 0.0136 & y & 46L & 0.15 & 40.79 & 41 & 4.62 & 23.324 & $\dots$ & 9 \\ 
NGC 1068        & 0.0038 & y & 3L & 67.8 & 42.33 & 22 & 4.62 & $>$25 & 1200$\pm$500 & 5  \\ 
NGC 1358       & 0.0134 & n & 23  & 0.18 & 40.86 & 43 & 0.86 & 23.6 & $\dots$ & 47 \\
NGC 1386       & 0.0029 & n & 23 & 10.2   & 41.27 & 38 & 0.27 & $>$24.342 & 1800$^{+400}_{-300}$ & 8 \\
NGC 1667       & 0.0152 & n & 15L,23 & 2.03 & 42.02 & 27   & 0.1 & $>$24 & 600 & 14 \\
NGC 1685       & 0.0152 & n & 23 & 9.09 & 42.67 & 28 & $<$2 & $\dots$ & $\dots$    & 39 \\ 
NGC 2273       & 0.0061 & y & 23K & 1.64 & 41.13 & 48 & 0.69 & $>$24.126 & 2200$_{-300}^{+400}$ & 8 \\ 
NGC 3081       & 0.0079 & y & 23K  & 1.95 & 41.43 & 27 & 1.3  & 23.819 & 610$_{-210}^{+390}$ & 49 \\
NGC 3281       & 0.0115 & n & 23 & 1.0 & 41.47 & 64 & 2.9 & 24.197 & 1180$_{-361}^{+400}$ & 50 \\
NGC 3982       & 0.0037 & n & 15L,23 & 0.66 & 40.3 & 15,31  & 0.057  & $>$ 24 & 6310$^{+3500}_{-3170}$ & 13 \\
NGC 4117       & 0.0031 & n & 23 & $\dots$ & $\dots$ & $\dots$  &  $<$23.2   & $\dots$ & $\dots$ & 39 \\
NGC 4388        & 0.0084 & y & 1A & 4.51 & 41.85 & 22 & 7.62 & 23.43 & 440$_{-90}^{+90}$ & 5 \\
NGC 4507       & 0.0118 & y & 23K & 4.98 & 42.19 & 43 & 12.8 & 23.643 & 117 & 51 \\ 
NGC 4941       & 0.0037 & n & 23  & 4.57 & 41.14 & 38 & 0.66     & 23.653  & 1600$_{-900}^{+700}$ & 40 \\ 
NGC 5135        & 0.0137 & n & 2,33A & 6.61  & 42.44 & 27 & 0.16 & $>$23.954 & 1700$_{-800}^{+600}$ & 8 \\
NGC 5283       & 0.0104 & n &15L,23 & 0.4  & 40.98 & 22 & 1.46  & 23.176  &  $<$220 & 7 \\
NGC 5728     & 0.0094 & n & 23,1A & 7.61 & 42.18 & 27 & 1.33 & 23.89 & 1100$^{+320}_{-270}$ & 13 \\
NGC 5643       & 0.004 & n & 23  & 6.62 & 41.37 & 43 & 0.84 & 23.845 &  500 & 52 \\ 
NGC 5347$^*$       & 0.0078 & y & 42K & 1.14 & 41.19 & 15 & 0.22  & $>$24 & 1300$\pm$500 & 63 \\NGC 5695       & 0.014  & n & 15L,23  & 0.081 & 40.55 & 22 & $<$0.01 & $\dots$  & $\dots$ & 39 \\
NGC 5929$^*$  & 0.0083 & y & 42K & 1.53 & 41.40 & 2 & 1.35 & 22.629 & $\dots$ & 9\\
NGC 6890       & 0.0081 & n & 23  & 0.5  & 40.86 & 43 & $\dots$  & $\dots$     & $\dots$ & $\dots$ \\ 
NGC 7672       & 0.0134 & n & 20L & $\dots$ & $\dots$   & $\dots$     & 28.6 & $\dots$ & $\dots$     & 39 \\
\\
Young et al. (1996)'s sample: \\ \\
IRAS 00521-7054 & 0.0689 & n & 1A & 0.36 & 42.62 & 1 &  $<$31.8 & $\dots$ & $\dots$  & 39 \\
IRAS 04103-2838 & 0.118 & n & 1A & $\dots$ & $\dots$ & $\dots$ & $\dots$ & $\dots$ & $\dots$ & $\dots$ \\
IRAS 04210+0400 & 0.046 & n & 1A & 0.554 & 42.44 & 1 & $\dots$  & $\dots$  & $\dots$ & $\dots$ \\
IRAS 04229-2528 & 0.044 & n & 1A  & 0.216 & 41.99 & 1  & $\dots$  & $\dots$  & $\dots$ & $\dots$ \\
IRAS 04385-0828 & 0.0151 & y & 15LP & 0.086 & 40.64 & 32 & 2.4  & $\dots$ & $\dots$ & 39 \\
IRAS 05189-2524 & 0.0426 & y  & 1A & 1.3  & 42.74 & 2  & 4.3  & 22.756 & 30$^{+50}_{-30}$ & 2  \\
IRAS 11058-1131 & 0.055 & y & 1A & 0.394 & 42.45 & 1 & 0.39 & $>$24 & 900 &  18 \\
MCG -3-34-64    & 0.0165 & y & 1A & 4.0 & 42.39 & 2  & 4.0 & 23.614 &  356$_{-143}^{+186}$ & 67 \\
IRAS 08277-0242 & 0.041 & n & 1A & 1.42 & 42.75 & 16   & $\dots$  & $\dots$   & $\dots$ & $\dots$ \\
IRAS 13452-4155 & 0.039 & n & 1A  &  $\dots$ & $\dots$     & $\dots$       & $\dots$  & $\dots$     & $\dots$   & $\dots$ \\
ESO 273-IG04    & 0.039 & y & 1A  &0.85  & 42.48 & 1  & $\dots$    & $\dots$    & $\dots$ & $\dots$ \\
IRAS 15480-0344 & 0.03 & y & 15P,1A & 5.03 & 43.02 & 16 & 0.37 & $>$24.204 & $<$2400 & 7 \\
IRAS 20210+1121 & 0.056 & n & 1A & 2.73 & 43.31 & 1 & 0.29 & $>$25 & 1650 &  18 \\ 
IRAS 20460+1925 & 0.181 & y & 1A   & 0.112 & 43.02 & 1 & 1.5 & 22.398 & $260_{-137}^{+145}$    & 49 \\
IRAS 22017+0319 & 0.0611 & y  & 15P,1A & 0.42 & 42.58 & 1 & 3.6  & 22.69 & $380^{+180}_{-160}$ & 18 \\
IRAS 23128-5919 & 0.045 & n & 1A & 0.101 & 41.68 & 1 & 0.13 & 22.681 & $\dots$ & 53 \\ 
NGC 5506      & 0.0062 & n & 15 & 3.33 & 41.45 & 58 & 58  & 22.46 & 86$_{-10}^{+24}$ & 11 \\
Mrk 463E        & 0.051  & y & 20L,1A & 1.25 & 42.89 &  22 & 1.46 & 23.51 & $340_{-90}^{+70}$ & 6 \\ 
NGC 4388        & 0.0084 & y & 1A & 4.51 & 41.85 & 22 & 7.62 & 23.43 & 440$_{-90}^{+90}$ & 5 \\
NGC 5252        & 0.023 & y & 1A & 0.921 & 42.05 & 1 & 10.7 & 22.461 & 44$\pm$28 & 13 \\
NGC 5728     & 0.0094 & n & 23,1A & 7.61 & 42.18 & 27 & 1.33 & 23.89 & 1100$^{+320}_{-270}$ & 13 \\
NGC 7496        & 0.005 & n & 1A & 0.3 & 40.22 & 1  & $<$8 & 22.699 & $\dots$ &  39 \\
NGC 7674        & 0.0289 & y & 1A,20L & 1.93 & 42.57 & 22 & 0.7 & $>$ 24 &  370$_{-170}^{+160}$ & 14 \\ \\
Other surveys: \\ \\
Mrk 1210        & 0.0135 & y & 54L & 5.8 & 42.37 & 55 & 9.3 & 23.263 & 108$^{+50}_{-65}$ &  56 \\ 
IRAS 18325-5926  & 0.0202 & y & 58A & 1.68 & 42.19 & 58 & 10 & 22.31 & 242 &  30 \\
MCG -5-23-16   & 0.008 & y & 58A & 4.09 & 41.81 & 49  & 70 & 22.25  & 35.2$^{+9.6}_{-10}$ & 59 \\ 
Circinus       & 0.0014 & y & 29 & 19.1 & 40.92 & 17 & 14 & 24.633 & 2250$^{+260}_{-300}$ & 4 \\ 
Mrk 477         & 0.038 & y & 54L & 12.4 & 43.62 & 22 & 1.2 & $>$24 & 490$_{-200}^{+250}$ & 49 \\
NGC 2992        & 0.0077 & y & 58A & 1.49 & 41.3 & 58 & 4.5 & 21.84 & 514$\pm$190 & 49 \\
NGC 7314        & 0.0047 & y & 58A & 17.7 & 42.41 & 49 & 41.2 & 22.02 & 147$_{-109}^{+128}$ & 60 \\ 
NGC 6300         & 0.0037 & n & 58A & 3.2 & 40.99 & 58 & 21.6 & 23.342 &148$_{-18}^{+18}$ & 65 \\ 
NGC 7212        & 0.027 & y & 54L & 3.2  & 42.73 & 43 & 0.69 & $>$24.204 & 900$_{-300}^{+200}$ & 7 \\
NGC 7590         & 0.0053 & n & 33A  & 0.168 & 40.02 & 27 & 1.2 & $<$20.964 & $\dots$ &  49 \\
Was 49b         & 0.063  & y & 54L  & 33.8 & 42.51 & 57 & 0.63 & 22.799 & $620\pm250$ & 21 \\
\enddata
\tablecomments{The Sy2s with footnote $^*$ denote the PBL was detected in later spectropolarimetric observation. Telescope: C = CTIO (4m), P = Palomar (5m), K = Keck (10m), L = Lick (3m), S = Subaru (8.2m), E = ESO (3.6m), A = AAT (3.9m).}
\tablerefs{(1) Young et al. 1996; (2) Lumsden et al. 2001; (3) Antonucci et al. 1985; (4) Smith \& Wilson 2001; (5) Cappi et al. 2006; (6) Imanishi \& Terashima et al. 2004; (7) Guainazzi et al. 2005a; (8) Guainazzi et al. 2005b; (9) tartarus.gsfc.nasa.gov; (10) Inglis et al. 1993; (11) Bianchi et al. 2003; (12) Levenson et al. 2005; (13) This work; (14) Bianchi et al. 2005a; (15) Tran 2001; (16) de Grijp et al. 1992; (17) Oliva et al. 1994; (18) Ueno et al. 2000; (19) Colina et al. 1991; (20) Miller et al. 1990; (21) Awaki et al. 2000; (22) Dahari \& De Robertis 1988; (23) Moran et al. 2000; (24) Murayama et al. 1998; (25) Matt et al. 2003; (26) Tran 1995; (27) Storchi-Bergmann et al. 1995; (28) Cruz-Gonzalez et al. 1994; (29) Alexander et al. 2000; (30) Iwasawa et al. 2004; (31) Ho et al. 1997; (32) Tran 2003; (33) Heisler et al. 1997; (34) Veilleux et al. 1995; (35) Vaceli et al. 1997; (36) Prieto et al. 2002; (37) de Robertis \& Osterbrock 1986; (38) Storchi-Bergmann \& Pastoriza 1989; (39) Polletta et al. 1996; (40) Maiolino et al. 1998; (41) Whittle et al. 1992; (42) Moran et al. 2001; (43) Mulchaey et al. 1994; (44) Acker et al. 1991; (45) Bianchi et al. 2005b; (46) Kay \& Moran 1998; (47) Fraquelli et al. 2003; (48) Lonsdale et al. 1992; (49) Bassani et al. 1999; (50) Vignali \& Comastri 2002; (51) Matt et al. 2004; (52) Guainazzi et al. 2004; (53) Franceschini et al. 2003; (54) Tran et al. 1992; (55) Terlevich et al. 1991; (56) Masanori et al. 2004; (57) Moran et al. 1992; (58) Lumsden et al. 2004; (59) Dewangan et al. 2003; (60) Dewangan \& Griffiths 2005; (61) Kollatschny et al. 1983; (62) Risaliti et al. 2000; (63) Levenson et al. 2006; (64) Storchi-Bergmann et al. 1992; (65) Matsumoto et al. 2004; (66) Ruiz et al. 1994; (67) Dadina \& Cappi 2004; (68) Duc et al. 1997; (69) Pernechele et al. 2003; (70) Braito et al. 2003.}
\end{deluxetable}

\clearpage

\begin{deluxetable}{llrrrrrrrr}
\tabletypesize{\scriptsize}
\rotate
\tablecaption{Best-fit parameters for the X-ray spectral analysis }
\tablewidth{0pt}
\tablehead{\colhead{Name} & \colhead{ XMM/Chandra obs. date} & \colhead{N$_{\rm H}$} & \colhead{$\Gamma$} & \colhead{Center Energy} & \colhead{EW(Fe K)} & \colhead{$f_{\rm s}$(\%)} & \colhead{$C/dof$} & \colhead{F$_{2-10keV}$} & \colhead{T} \\
\hspace*{10.mm}(1) & (2) & (3) &(4) & (5) & (6) & (7) & (8) & (9) & (10)}
\startdata
NGC 34  
        & XMM 2002 Dec 22 & $>$100 & 2.38$\pm0.1$ & 6.4$^{\dag}$ & $<$321& $\dots$ & 362/448 & 0.23 & 0.03 \\
NGC 3982 & XMM 2004 Jun 15 & $>$100 & 3.74$^{+1.8}_{-1.6}$ & 6.4$^{\dag}$ & 6310$^{+3500}_{-3170}$ & $\dots$ & 109/135 & 0.057 & 0.086 \\
NGC 5728 & Chandra 2003 Jun 27 & 78$^{+15}_{-14}$ & 2.73$\pm$0.17 & 6.39$\pm0.03$ & 1100$^{+320}_{-270}$  &  0.4$\pm$0.03 & 299/300 & 1.33  & 0.17 \\
NGC 6552 & XMM 2002 Oct 18 & 71$^{+40}_{-10}$ & 2.8$^{+0.37}_{-0.13}$ & 6.4$^{\dag}$  & 1408$^{+668}_{-883}$ & 0.75$^{+0.09}_{-0.22}$ & 121/166 & 0.43 & 0.27 \\NGC 7172 & XMM 2002 Nov 18 & 8.7$\pm$0.57 & 1.49$\pm$0.13 & 6.4$^{\dag}$ & 40$\pm$30 & 0.54$\pm$0.1 & 463/495 & 22 & 550 \\
Mrk 573   & XMM 2004 Jan 15 & $>$100 & 3.7$^{+0.08}_{-0.10}$   & 6.4$^{\dag}$ & 2800$^{+1820}_{-1220}$ & $\dots$ & 394/399 & 0.12 & 0.07 \\
IRAS 01475-0740 & XMM 2004 Jan 21 & 0.39$^{+0.04}_{-0.02}$ & 2.06$\pm0.06$ & 6.4$^{\dag}$ & 130($<$344) & $\dots$ & 631/776 & 0.75 & 0.92 \\
Mrk 1066 & Chandra 2004 Jul 18 &  $>$100 & 2.75$^{+0.17}_{-0.07}$ & 6.34$^{+0.17}_{-0.06}$ & 1120$^{+850}_{-650}$ & $\dots$  & 497/505 & 0.23 &  0.05 \\
\enddata
\tablecomments{ Col. (1): galaxy name. Col. (2):  Observation date.  Col. (3) power-law photon index. Col. (4): measured absorption column density, in units of 10$^{22}$ cm$^{-2}$. Col. (5): the Fe line energy in units of keV. Col. (6): the Fe line equivalent width  in units of eV. Col. (7): scattering fraction of the soft component. Col. (8): C statistic and number of degrees of freedom (dof). Col (9): the fitted 2-10 Kev flux in units of 10$^{-12}$ erg s$^{-1}$ cm$^{-2}$. Col (10) T: the ratio of F$_{\rm 2-10~keV}$/F$_{\rm [O~III]}$. $^{\dag}$ : fixed.}
\end{deluxetable}

\clearpage

\begin{deluxetable}{lcclr}
\tabletypesize{\scriptsize}
\tablecaption{A summary of the Statistical Properties for Sy2s}
\tablewidth{0pt}
\tablehead{
\colhead{Parameters} & \colhead{Sy2s with PBL} & \colhead{Sy2s without PBL}  & \colhead{Note} & \colhead{p$_{\rm null}^a$}    }
\startdata
log(T) & -0.087$\pm$0.145 & -0.342$\pm$0.217 & total & 25\% \\
log(T) & -0.201$\pm$0.138 & -0.87$\pm$0.186 & luminous & 0.9\% \\
log(T)$^b$ & -0.093$\pm$0.144 & -0.436$\pm$0.199 & total & 13.5\% \\
log(T)$^b$ &  -0.203$\pm$ 0.137 & -0.899$\pm$0.179 & luminous & 0.68\% \\
log[EW(Fe)]           & 2.654$\pm$0.107 & 2.85$\pm$0.127 & total & 18.6\% \\
log[EW(Fe)]           & 2.626$\pm$0.107 & 2.999$\pm$0.066 & luminous & 4.67\% \\log(N$_{\rm H}$)            & 23.755$\pm$0.190 & 23.852$\pm$0.274 & total & 66.5\% \\
log(N$_{\rm H}$)            & 23.739$\pm$0.212 & 24.428$\pm$0.192 & luminous & 7.7\% \\
log(L$_{\rm [O~III]}$) & 42.099$\pm$0.13 & 41.338$\pm$0.188 & total & 0.34\% \\
log(L$_{\rm [O~III]}$) & 42.198$\pm$0.121 & 41.991$\pm$0.157 & luminous & 30.6\% \\
log(z)                  & 0.020$\pm$0.003 & 0.017$\pm$0.004 & luminous & 29.4\% \\
\enddata
\tablecomments{ $^a$ the possibility p$_{\rm null}$ is for the null hypothesis 
that the two distributions are drawn at random from the same parent population.
$^b$ including 8 hard X-ray upper limits stated in \S2.
When there are censored data, we use Gehan's generalized Wilconxon test-permutation variance (GGW test, one kind of ASURV test). "total": all Sy2s with X-ray data in the sample, "luminous": for Sy2s with L$_{\rm [O~III]}$ $>$ 10$^{41}$ erg s$^{-1}$ only.
}
\end{deluxetable}


\begin{thebibliography}{}
\bibitem[1991]{ack91} Acker, A., Stenholm, B., \& Veron, P. 1991, A\&AS, 87, 499
\bibitem[Alexander(2001)]{ale01} Alexander, D. M. 2001, \mnras, 320, L15
\bibitem[2000]{ale00} Alexander, D. M., Heisler, C. A., Young, S., et al. 2000, \mnras, 313, 815
\bibitem[1985]{Ant85} Antonucci, R., \& Miller, J. S. 1985, \apj, 297, 621
\bibitem[Antonucci(1993)]{ant93} Antonucci, R. 1993, ARA\&A, 31, 473
\bibitem[Awaki et al. (2000)]{awa00} Awaki, H., Ueno, S., Taniguchi, Y., \& Weaver, K. A. 2000, \apj, 542, 175
\bibitem[Bassani et al. (1999)]{bas99} Bassani, L., Dadina, M., Maiolino, R., Salvati, M., Risaliti, G., della Ceca, R., Matt, G., \& Zamorani, G. 1999, \apjs, 121, 473
\bibitem[2003]{Bianchi03} Bianchi, S., Balestra, I., Matt, G., et al. 2003, \aap, 402, 141
\bibitem[2005a]{Bianchi05a} Bianchi, S., Miniutti, G., Fabian, A., \& Iwasawa, K. 2005a, \mnras, 360, 380
\bibitem[2005b]{Bianchi05b} Bianchi, S., Guainaazi, M., Matt, G., Chiaberge, M., Iwasawa, K., Fiore, F., \& Maiolino, R. 2005b, \aap, 442, 185
\bibitem[2003]{Bai03} Braito, V., Framcescjomo, A., Della Ceca,R., et al. 2003, \aap, 398, 107
\bibitem[1979]{cas79} Cash. W. 1979, \apj, 228, 939
\bibitem[2005]{Cappi} Cappi, M., Panessa, F., Bassani, L., et al. 2006, \aap, 446, 459
\bibitem[2002]{che02} Cheng, L. P., Zhao, Y. H., \& Wei, J. Y. 2002, CHJAA, 2, 408
\bibitem[1991]{Col91} Colina L., Sparks W. B., \& Macchetto F., 1991, \apj, 370, 102
\bibitem[1994]{cru94} Cruz-Gonzalez, I., et al. 1994, \apjs, 94, 47
\bibitem[2004]{dadina} Dadina, M., \& Cappi, M. 2004, \aap, 413, 921
\bibitem[1992]{deG92} de Grijp, M. H. K., Keel, W. C., Miley, G. K., \& Lub, J. 1992, A\&AS, 96, 389
\bibitem[1988]{dah88} Dahari, O., \& De Robertis, M. M. 1988, \apjs, 67, 249
\bibitem[1986]{der86} de Robertis, M. M., \& Osterbrock. D. E., 1986, \apj, 301, 98
\bibitem[2003]{Dewangan03} Dewangan, G. C., Griffiths, R. E., \& Schurch, N. J. 2003, ApJ, 592, 52
\bibitem[2005]{Dewangan05} Dewangan, G. C., \& Griffiths, R. E., 2005, ApJ, 625, L31
\bibitem[1997]{Duc97} Duc, P. A., Mirabel, I. F., \& Maza, J. 1997, A\&AS, 124, 533
\bibitem[Feigelson \& Nelson(1985)]{fei85} Feigelson, E. D., \& Nelson, P. I. 1985, \apj, 293, 192
\bibitem[2003]{Franceschini} Franceschini, A., et al. 2003, \mnras, 343, 1181
\bibitem[2003]{fra03} Fraquelli, H. A., Storchi-Bergmann, T., \& Levenson, N. A. 2003, \mnras, 341, 449 
\bibitem[Georgantopoulos(2003)]{geo03} Georgantopoulos, I., \& Zezas, A. 2003, \apj, 594, 704 
\bibitem[1998]{Guainazzi98} Guainazzi, M., Matt, G., Antonelli, L. A., Fiore, F., Piro, L. \& Ueno, S. 1998, \mnras, 298, 842
\bibitem[2004] {Guainazzi04} Guainazzi, M., Rodriguez-Pascual. P., Fabian, A. C., Iwasawa, K., \& Matt, G. 2004, \mnras, 355, 297
\bibitem[2005a]{Guainazzi05a} Guainazzi, M., Fabian, A. C., Iwasawa, K., Matt, G., Fiore, F. 2005a, \mnras, 356, 295
\bibitem[2005b]{Guainazzi05b} Guainazzi, M., Matt, G., \& Perola, G. C. 2005b, \aap, 444, 119
\bibitem[Gu et al. (2001)]{gu01} Gu, Q., Maiolino, R., \& Dultzin-Hacyan, D. 2001, \aap, 366, 765
\bibitem[Gu \& Huang(2002)]{gu02} Gu, Q., \& Huang, J. 2002, \apj, 579, 205
\bibitem[Heisler et al.(1997)]{hei97} Heisler, C. A., Lumsden, S. L., \& Bailey, J. A. 1997, Nature, 385, 700
\bibitem[1997]{Ho97} Ho, L., Filippenko, V., \& Sargent, W. L. W., 1997, \apjs, 112, 315
\bibitem[1993]{Ing93} Inglis, M. D., et al. 1993, \mnras, 263, 895
\bibitem[2004]{Imanishi} Imanishi, M., \& Terashima, Y. 2004, \aj, 127, 758
\bibitem[2004]{Iwasawa04} Iwasawa, K., Lee, J. C., \& Young, A. J. et al. 2004, \mnras, 347, 4111
\bibitem[1998]{kay98} Kay, L. E., \& Moran, E. C., 1998, PASP, 110, 1003 
\bibitem[1983] Kollatschny, W., Fricke, K. J., Biermann, P., et al. 1983, \aap, 119, 80
\bibitem[1992]{lon92} Lonsdale, C. J., 1992, \apj, 391, 629
\bibitem[Lumsden et al.(2001]{lum01} Lumsdem, S. L., Heisler, C. A., \& Bailey, J. A., Hough, J. H., Young, S. 2001, \mnras, 327, 459
\bibitem[Lumsden &Alexander2001]{luma01} Lumsdem, S. L. \& Alexander, D. M. 2001, MNRAS, 328, L32
\bibitem[Lumsden et al.(2004)]{lum04} Lumsden, S., Alexander, D., \& Hough, J., 2004, \mnras, 348, 1451
\bibitem[2005]{levenson} Levenson, N. A., Weaver, K. A., Heckman, T. M., Awaki, H., \& Terashima, Y. 2005, \apj, 618, 167
\bibitem[2006]{lev06} Levenson, N. A., Heckman, T. M., Krolik, J. H., et al. 2006, astro-ph/0605438
\bibitem[Maiolino(1998)]{mai98} Maiolino, R., Salvati, M., Bassani, L., Dadina, M., della Ceca, R., Matt, G., Risaliti, G., \& Zamorani, G. 1998, \aap, 338, 781
\bibitem[1995]{mag95} Magdziarz, P., \& Zdziarshi, A. A. 1995, \mnras, 273, 837
\bibitem[2004]{Masanori} Masanori, O., Yasushi, F., \& Naoko, I. 2004, PASJ, 56, 425
\bibitem[2004]{Matsumoto} Matsumoto, C., Nava, A., Maddox, L. A., Leighly, K. M., Grupe, D., Awaki, H., \& Ueno, S. 2004, \apj, 617, 930
\bibitem[Matt(2000)]{Matt00a}Matt, G., 2000a, \aap, 355, L31
\bibitem[Matt(2000)]{Matt00b}Matt, G., Fabian, A. C., Guainazzi, M., et al. 2000b, \mnras, 318, 173
\bibitem[2003]{matt03} Matt, G., Bianchi, S., Guainazzi, M., Brandt, W. N., Fabian, A. C., Iwasawa, K., \& Perola, G. C. 2003, \aap, 339, 519
\bibitem[2004]{matt04} Matt, G., Bianchi, S., D'Ammando, F., \& Martocchia, A. 2004, \aap, 421, 473
\bibitem[1990]{mil90} Miller, J. S., \& Goodrich, R. W., 1990, \apj, 355, 456
\bibitem[Moran(1992)]{mor92} Moran, E. C., Halpern, J. P., Bothun, G. D., \& Becker, R. H. 1992, AJ, 104, 990
\bibitem[Moran(2000)]{mor00} Moran, E. C., Barth, A. J., Kay, L. E., \& Filippenko, A. V. 2000, \apj, 540, L73
\bibitem[2001]{mor01} Moran, E. C., Kay, L. E., Davis, M., Filippenko, A. V., \& Barth, A. J. 2001, \apj, 556, L75
\bibitem[1994]{mul} Mulchaey, J., Koratkar, A., Ward, M. J., et al. 1994, \apj, 436, 586
\bibitem[1998]{Mur98} Murayama, T., Taniguchi, Y., \& Iwasawa, K. 1998, \aj, 115, 460
\bibitem[Nicastro03]{nic03} Nicastro, F. Martocchia, A. \& Matt, G. 2003, ApJ, 589, L13
\bibitem[1994]{oli94} Oliva, E., Salvati, M., Moorwood, A. F. M., \& Marconi, A. 1994, \aap, 288, 457
\bibitem[Panessa \& Bassani(2002)]{pan02} Panessa, F., \& Bassani, L. 2002, \aap, 394, 435
\bibitem[1996]{Polletta} Polletta, M., Bassani, L., Malaguti, G., Palumbo, G. G. C., \& Caroli, E. 1996, \apjs, 106, 399
\bibitem[2002]{pri02} Prieto, M. Almudena, Perez Garcia, A. M., \& Rodriguez Espinosa, J. M. 2002, \mnras, 329, 309
\bibitem[2003]{Per03} Pernechele, C., Berta, S., Marconi, A. 2003, \mnras, 338, 13
\bibitem[2000]{Risaliti} Risaliti, G., Gilli, R., Maiolino, R., Salvati, M., 2000, \aap, 357, 13
\bibitem[Risaliti(2002)]{ris02} Risaliti, G. 2002, \aap, 386, 379
\bibitem[1994[{Rui94} Ruiz, M., Rieke, G. H., \& Schmidt, G. D. 1994, \apj, 423, 608
\bibitem[2001]{Smith} Smith, D. A., \& Wilson, A. S., 2001, \apj, 557, 180
\bibitem[1989]{SB89} Storchi-Bergmann, T., \& Pastoriza, M. G., 1989, \apj, 347, 195
\bibitem[1992]{SB92} Storchi-Bergmann, T., Wilson, A. S., Baldwin, J. A. 1992, \apj, 396, 45
\bibitem[1995]{SB95} Storchi-Bergmann, T., Kinney, A. L., \& Challis, P. 1995, \apjs, 98, 103
\bibitem[1991]{ter91} Terlevich, R., Melnick, J., Masegosa, J., Moles, M., \& Copetti, M., 1991, A\&AS, 91, 285
\bibitem[1997]{turner} Turner, T. J., George, I. M., Nandra, K., Mushotzky, \& R. F. 1997, \apj, 488, 164
\bibitem[1992]{tra92} Tran, H. D., Miller, J. S., \& Kay, L. E., 1992, \apj, 397, 452
\bibitem[Tran(1995)]{tra95} Tran, H. D., 1995, \apj, 440, 565
\bibitem[Tran (2001)]{tra01} Tran, H. D., 2001, \apj, 554, L19
\bibitem[Tran(2003)]{tra03} Tran, H. D., 2003, \apj, 583, 632
\bibitem[2000]{ueno} Ueno, S., Ward, M. J., O'Brien, P. T., et al. 2000, AdSpR, 25, 823 
\bibitem[1997]{vac} Vaceli, M. S., Viegas, S. M., Gruenwald, R., \& Souza, R. E. 1997, \aj, 114, 1345
\bibitem[1995]{Veilleux95} Veilleux, S., Kim, D. C., Sanders, D., Mazzarella, J. M., \& Soifer, B. T., 1995, \apjs, 98, 171
\bibitem[1997]{Veilleux97} Veilleux, S., Goodrich, R. W., Hill, G. J., 1997, \apj, 477, 631
\bibitem[2002]{vignali02} Vignali, C., \& Comastri, A. 2002, \aap, 381, 834
\bibitem[2004]{vignali04} Vignali, C., Alexander, D. M., \& Comastri, A. 2004, \mnras, 354, 720
\bibitem[1992]{wit92} Whittle, M., 1992, \apjs, 79, 49
\bibitem[1996]{Young} Young, S., Hough, J. H., Efstathiou, A., Wills, B. J., Bailey, J. A., Ward, M. J., \& Axon, D. J. 1996, \mnras, 281, 1206
\bibitem[2005]{Yu}Yu, P., \& Hwang, C. 2005, ApJ, 631, 720
\end{thebibliography}
\end{document}